**Combined Ab-Initio and Empirical Model of the thermal conductivity of Uranium, Uranium-Zirconium, and Uranium-Molybdenum**

Shuxiang Zhou[1], Ryan Jacobs[1], Wei Xie[1], Eric Tea[2], Celine Hin[2,3], Dane Morgan[1,*]

[1]Department of Materials Science and Engineering, University of Wisconsin-Madison, Madison, WI 53706, USA

[2]Department of Mechanical Engineering, Virginia Polytechnic Institute and State University, Blacksburg, VA 24061, USA

[3]Department of Materials Science and Engineering, Virginia Polytechnic Institute and State University, Blacksburg, VA 24061, USA

*Corresponding author: Dane Morgan, e-mail: ddmorgan@wisc.edu

## Abstract

In this work, we developed a practical and general modeling approach for thermal conductivity of metals and metal alloys that integrates ab-initio and semi-empirical physics-based models to maximize the strengths of both techniques. The approach supports creation of highly accurate, mechanistic, and extensible thermal conductivity modeling of alloys. The model was demonstrated on α-U and U-rich U-Zr and U-Mo alloys, which are potential fuels for advanced nuclear reactors. The safe use of U-based fuels requires quantitative understanding of thermal transport characteristics of the fuel. The model incorporated both phonon and electron contributions, displayed good agreement with experimental data over a wide temperature range, and provided insight into the different physical factors that govern the thermal conductivity under different temperatures. This model is general enough to incorporate more complex effects like additional alloying species, defects, transmutation products and noble gas bubbles to predict the behavior of complex metallic alloys like U-alloy fuel systems under burnup.

## 1    Introduction

Thermal conductivity describes the rate at which a material transfers heat, and high-quality thermal conductivity data is critical to many technologies, ranging from thermoelectrics to nuclear reactors. However, due to the high cost and low efficiency of traditional thermal conductivity measurement techniques [1,2], experimental thermal conductivity data is limited and often difficult to obtain over a wide temperature and composition range. Furthermore, an understanding of the origins of measured thermal conductivity in terms of different scattering mechanisms is important for predicting how thermal conductivity might change over time or be controlled through rational materials design. Thermal conductivity modeling is a powerful tool to both understand the impact of different factors on thermal conductivity and interpolate and extrapolate existing data.

In metals and metal alloys, the conduction of heat is controlled by interactions between phonons, electrons, and defects, including phonon-phonon, electron-phonon, electron-electron, and phonon/electron-defect scattering. Many models for different contributions to thermal conductivity exist, but each bring different strengths and weaknesses. On one extreme are simple empirical interpolation formulae [3,4], which are effective at quantitatively interpolating measured data but provide little physical insight, making it difficult to utilize them for guidance in designing improved materials. These models also require significant measured data for fitting and may have large errors extrapolating outside the fitted data range due to their lack of full mechanistic description. On the other extreme are highly complex ab-initio based simulations, which predict thermal conductivity contributions from the fundamental equations of quantum mechanics and heat transport [5,6]. These methods require little or no empirical data and provide detailed mechanistic information and insight, but are often very technically challenging to implement, limited to a specific mechanism and/or composition or temperature range, and of limited quantitative accuracy. There are also modeling approaches for different scattering mechanisms intermediate to these extremes that integrate a significant amount of physics and empirical fitting, at varying levels. For modeling of technologically important alloy systems, where typically some thermal and electrical conductivity data is available, quantitative prediction is required, and materials optimization is often a goal. However, it has been unclear how to integrate the available approaches most effectively for such cases. To solve this problem, we developed in this work a model for metal alloy thermal conductivity that includes all of the different electron and phonon contributions and their scattering mechanisms using a practical and accurate combination of ab-initio methods and empirical fitting. This approaches leverages new integrations of 6 key component models into a complete model for metal thermal conductivity: (i) DFT phonon thermal conductivity calculation, (ii) the Wiedemann-Franz law for electronic contributions to thermal conductivity, (iii) electron-electron scattering from DFT band structure and Boltzmann Transport Equation (BTE) (including an empirical temperature-dependent relaxation time), (iv) electron-phonon scattering from semi-classical models using the DFT-calculated phonon spectrum, (v) Nordheim-type models for residual, impurity, and alloying effects, (vi) the effects of resistivity saturation. Taken together, these component models provide a significantly more complete physics-based model of thermal conductivity for metals than has been available previously. The integration also allows greater accuracy when fitting to limited data, fewer fitting parameters, and more accurate physics in some of the component models.

The specific systems for which we have demonstrated this modeling approach are metallic uranium (U) alloys. Metallic U alloys possess multiple advantages compared to the $UO_2$-based nuclear fuels that are widely used in both current thermal and fast nuclear reactors, including higher thermal conductivity, high burn-up capability, good transient overpower capability, ease of recycling, and lower radiotoxicity of nuclear waste [7,8]. These advantages make metallic U alloys promising materials for use as nuclear fuels in thermal reactors and especially in the future deployment of fast reactors as Generation IV nuclear reactor designs [9]. While metallic U alloys have higher thermal conductivities than $UO_2$-based fuels, they also have much lower melting temperatures. Consequently, temperature control of reactors containing metallic U alloy fuels becomes a critically important issue. Thus, the thermal conductivity of metallic U alloys is an essential property influenced by many factors, including alloying, impurities, and point and extended defects. Developing a physical understanding of the thermal conductivity of metallic U alloy fuels will help improve temperature control and guide its use in reactor environments. Unfortunately, no such general integrated approach has been developed in the literature.

Since 1950, the thermal conductivity of α-U has been measured in numerous experiments. In particular, the experimental data available before 1970 has been summarized by Touloukian et al. [10]. Since 1970, the main experimental data has been reported by Takahashi et al. [11], Hall et al. [12], and Kaity et al. [13]. While there is extensive experimental data on the thermal conductivity of α-U, a physics-based computational model that includes the fundamental factors contributing to the thermal conductivity of α-U under different temperature conditions has not been established. Therefore, we here developed an approach that integrates what can be determined accurately with state of the art ab-initio methods with physics-based functional forms for fitting, and demonstrated this approach to establish a full physics-based model for α-U thermal conductivity. We then extended the model to U-rich U-Mo and U-Zr alloys.

The goal of the present study is to develop a physics-based computational model of thermal conductivity in metals and metal alloys. The model is demonstrated on α-U and binary U-rich alloys containing Zr and Mo as a foundation for understanding the more complex thermal conductivity of realistic U alloy fuels. Here, we use all known experimental data of thermal conductivity and electrical resistivity in α-U combined with a density functional theory (DFT)-based computational framework to construct a model of the phonon and electron contributions to the thermal conductivity in pure α-U. This physics-based computational model provides an understanding of the dominant mechanisms for thermal conductivity and, importantly, is extensible to more realistic metallic alloy fuels that include physical effects resulting from alloying elements, impurities, transmutation products, radiation-induced defects and noble gas bubbles.

## 2    Methods

### 2.1    Computational model of thermal conductivity in pure α-U and U-rich alloys

#### 2.1.1 Model for pure α-U

In metals like α-U, the thermal conductivity $\kappa$ is the sum of electron and phonon thermal conductivities [14]:

$$\kappa = \kappa_e + \kappa_{ph} \,, \tag{1}$$

where $\kappa_e$ is controlled by electron-electron, electron-phonon and electron-defect scattering, the relative strengths of which are material- and environment-dependent, and $\kappa_{ph}$ is controlled by phonon-phonon, phonon-grain boundary and phonon-defect scattering. A physics-based computational model that separately includes electron and phonon contributions and their relative scattering mechanisms cannot be obtained only using the available experimental data. Therefore, DFT calculations were used to calculate the structural, electronic and vibrational properties of α-U to analyze the phonon scatterings and electron-electron scattering processes. We combined these results with semi-empirical analytic models of electron-phonon, phonon-defect and phonon-grain boundary scattering fitted to available experimental electrical resistivity data of α-U to form our full model. In the following paragraphs, we describe how each term is modeled.

The phonon thermal conductivity $\kappa_{\text{ph}}$ can be calculated using the phonon Boltzmann transport equation (BTE) given by [15]:

$$\boldsymbol{\kappa}_{ph} = \frac{1}{3\Omega N_k} \sum_{k\lambda} c_{k\lambda} \boldsymbol{v}_{k\lambda} \otimes \boldsymbol{v}_{k\lambda} \tau_{k\lambda}^{ph} \,, \tag{2}$$

where $k$ represents the wavevector, $\lambda$ represents the different phonon branches, $N$ is the total number of discrete k-points, and $\Omega$ is the volume of the unit cell. The quantities inside the summation consist of $c_{k\lambda}$, $\boldsymbol{v}_{k\lambda}$, and $\tau_{k\lambda}^{ph}$, which are the heat capacity, phonon group velocity and phonon relaxation time for each wavevector and phonon branch, respectively. The $c_{k\lambda}$ values are evaluated from the phonon density of states using the Bose-Einstein distribution. The $\boldsymbol{v}_{k\lambda}$ values are obtained by calculating the gradient of the phonon dispersion relation. Considering different phonon scattering contributions, the phonon relaxation time $\tau_{k\lambda}^{ph}$ for each wavevector $k$ and phonon branch $\lambda$ can be divided into three contributions, one each for phonon-phonon, phonon-grain boundary and phonon-defect scattering, using Matthiessen's rule [16]:

$$\frac{1}{\tau_{k\lambda}^{ph}} = \frac{1}{\tau_{k\lambda}^{ph-ph}} + \frac{1}{\tau_{k\lambda}^{ph-gb}} + \frac{1}{\tau_{k\lambda}^{ph-def}}, \tag{3}$$

where $\tau_{k\lambda}^{ph-ph}$, $\tau_{k\lambda}^{ph-gb}$ and $\tau_{k\lambda}^{ph-def}$ are the relaxation time of phonon-phonon, phonon-grain boundary and phonon defect scattering contribution, respectively. There are other phonon scattering processes that are neglected in our model, specifically, phonon-electron, phonon-dislocation and phonon-isotope scattering. They are discussed in Section 4.1.

The phonon relaxation time for phonon-phonon scattering $\tau_{ph-ph}$ is calculated from Fermi's Golden rule using the harmonic and third-order anharmonic force constants from DFT calculations [6]. The phonon relaxation time for phonon-grain boundary scattering $\tau_{ph-gb}$ can be estimated by [16,17]:

$$\tau_{k\lambda}^{ph-gb} = \frac{L}{v_{k\lambda}},$$ (4)

where $L$ is the grain size and $v_{k\lambda}$ is the phonon group velocity. We directly evaluate the value of $L$ by experimental data and in this work use a grain size of ~0.015mm in diameter [18]. The phonon relaxation time for phonon-defect scattering can be approximated by [16,17]:

$$\tau_{k\lambda}^{ph-def^{-1}} = A\omega_{k\lambda}^4,$$ (5)

where $\omega$ is the phonon frequency and $A$ is a constant. In our model $A$ is assumed to be an isotropic fitting parameter obtained from fitting to low temperature thermal conductivity data. Thus, calculating $\kappa_{ph}$ requires calculation of the harmonic and third-order anharmonic force constants and the associated phonon dispersion relations and couplings. All of these phonon-related quantities can be determined from DFT calculations using the Vienna Ab initio Simulation Package (VASP) code [19,20]. This approach for phonon-phonon scattering has been successfully applied to calculate the phonon contribution to the thermal conductivity in a number of systems, such as PbTe and PbSe materials [21]. To summarize, $\kappa_{ph}$ is calculated using Eq. (2), where the contributions of different phonon scatterings are combined using Eq. (3).

The electronic thermal conductivity $\kappa_e$ can be evaluated from electrical resistivity using the Wiedemann-Franz law [16]:

$$\kappa_e = \frac{\pi^2}{3}\left(\frac{k_B}{e}\right)^2 \frac{T}{\rho},$$ (6)

where $T$ is the temperature and $\rho$ is the electrical resistivity. It is useful to treat the resistivity as having an ideal contribution, which is modeled accurately with the semi-classical BTE approach, and then modify this ideal contribution with a saturation effect, which occurs in some metals like U due to the breakdown of the semi-classical approach for electrons with wavelengths approaching the length of interatomic separations [22]. We will take this approach in the present work, and first consider the ideal contribution, $\rho_{id}$, to the electrical resistivity. $\rho_{id}$ can be divided into two scattering contributions using Matthiessen's rule [16]:

$$\rho_{id} = \rho_{e-e} + \rho_{e-ph},$$ (7)

where $\rho_{e-e}$ is the electron-electron scattering contribution and $\rho_{e-ph}$ is the electron-phonon scattering contribution. $\rho_{e-e}$ can be obtained from the electrical conductivity tensor $\sigma$, total electronic relaxation time $\tau$, and the electronic relaxation time for electron-electron scattering $\tau_{e-e}$:

$$\rho_{e-e} = \frac{1}{\sigma_{e-e}} = \frac{1}{\frac{\sigma}{\tau}\tau_{e-e}}.$$ (8)

$\left(\frac{\sigma}{\tau}\right)$ is calculated using the BTE with the relaxation time approximation (RTA) and the rigid band approximation as follows [23]. The full tensor of $\sigma$ can be calculated from the conductivity distributions:

$$\sigma_{\alpha\beta} = \frac{1}{\Omega}\int \sigma_{\alpha\beta}(\varepsilon)\left[-\frac{\partial f_\mu(T;\varepsilon)}{\partial \varepsilon}\right]d\varepsilon ,$$ (9)

where $\sigma_{\alpha\beta}$ is the full tensor that we denote as just $\sigma$ in Eq. (8) (and will denote as just $\sigma$ throughout this paper), $f$ is the Fermi-Dirac distribution function, $\Omega$ is the volume system, $T$ is the temperature, $\varepsilon$ is the band energy, $\mu$ is the Fermi level, and $\sigma_{\alpha\beta}(\varepsilon)$ is the transport distribution given by:

$$\sigma_{\alpha\beta}(\varepsilon) = \frac{1}{N}\sum_{i,k}\sigma_{\alpha\beta}(i,k)\frac{\delta(\varepsilon-\varepsilon_{i,k})}{d\varepsilon}, \tag{10}$$

where $k$ represents elements of a set of k-points in reciprocal space, $N$ is the number of k-points sampled, and $i$ is the band index. $\sigma_{\alpha\beta}(i,k)$ is the conductivity tensor, which depends on the electron group velocity $v_\alpha$, the elementary charge $e$, and the relaxation times $\tau_{i,k}$:

$$\sigma_{\alpha\beta}(i,k) = e^2\tau_{i,k}v_\alpha(i,k)v_\beta(i,k). \tag{11}$$

Finally, $\frac{\partial f_\mu(T;\varepsilon)}{\partial\varepsilon}$ is the temperature smearing, which is determined using the electronic density of states and the Fermi-Dirac distribution evaluated at the appropriate temperature. Therefore, the input parameters for calculating the electrical conductivity tensor $\sigma$ are the relaxation times $\tau_{i,k}$ and the electronic band structure information calculated from DFT. Here, we simplify the input of $\tau_{i,k}$ by assuming it is a constant which neither depends on band index $i$ nor direction $k$, which has been shown to be a satisfactory assumption for many metals [24,25]. With this approximation, $(\frac{\sigma}{\tau})$ depends on only the electronic band structure through Eqs. (9-11) and its values can be obtained from DFT-BTE calculations using VASP for the band structure and the BoltzTrap code [23]. In metals, the electron-electron scattering relaxation time $\tau_{e-e}$ is dependent on temperature $T$ and can be approximated by:

$$\tau_{e-e} = BT^{-2}, \tag{12}$$

where $B$ is typically assumed to be a constant [26–28].

For the electron-phonon scattering contribution $\rho_{e-ph}$, we use semi-empirical models and fit the unknown terms using available experimental data. We note that $\rho_{e-ph}$ can be predicted from DFT calculations [5]. However, we did not pursue this path due to the limitations of present methods for modeling U at high temperature, which are discussed in Section 4.2, and instead fit semi-empirical models to the available experimental data and DFT-calculated phonon spectra. Following Ziman's approach [16], which assumes a spherical Fermi surface and the deformation potential approach, $\rho_{e-ph}$ can be written

$$\rho_{e-ph} \propto \frac{1}{T} \int_0^R \frac{k^5}{(e^{\hbar\omega/k_B T}-1)(1-e^{-\hbar\omega/k_B T})} dk, \qquad (13)$$

where $k$ represents the wavevector, $\nu$ represents the phonon frequency, and $R$ is the Debye radius. By further assuming the Debye phonon spectrum, the Bloch-Gruneisen formula can be obtained from Eq. (13) [16]:

$$\rho_{e-ph} \propto \left(\frac{T}{\theta_D^R}\right)^5 \int_0^{\frac{\theta_D^R}{T}} \frac{x^5}{(e^x-1)(1-e^{-x})} dx, \qquad (14)$$

where $\theta_D^R$ is the Debye temperature obtained from resistivity measurements. We can improve the accuracy of this model by using the full phonon spectrum for α-U obtained by DFT in place of the approximate Debye phonon spectrum, which yields:

$$\rho_{e-ph} = \frac{C}{T} \sum_{k\lambda} \frac{k^3}{(e^{\hbar\omega_{k\lambda}/k_B T}-1)(1-e^{-\hbar\omega_{k\lambda}/k_B T})}, \qquad (15)$$

where $k$ represents the reduced wave vector in the first Brillouin zone, $\lambda$ represents the different phonon branches, $\omega$ represents the phonon frequency and $C$ is a constant which, in general, depends on crystallographic direction. The validity of using this method to calculate the electron-phonon scattering contribution on resistivity is discussed in Section 4.2.

Within the framework described above, the ideal resistivity $\rho_{id}$ given in Eq. (5) grows linearly with temperature $T$ at high temperature due to the dominance of electron-phonon scattering [22]. However, in some metals, including α-U (see Figure 6), the resistivity at high temperature is not a linear function of $T$ due to the resistivity saturation effect mentioned above [22]. We model the total resistivity $\rho$ in the presence of saturation using Wiesmann, et al.'s parallel resistor formula [29]:

$$\rho = (\rho_{id}^{-1} + \rho_{sat}^{-1})^{-1}, \tag{16}$$

where the saturation resistivity $\rho_{sat}$ is assumed to be a temperature-independent constant. This approach has been successfully applied to multiple A15 compounds, e.g. Nb$_3$Sn [29].

As the final component of our resistivity model, we include the electron-defect scattering for pure α-U, which is assumed to arise from point defects and dislocations, and to some extent the elemental impurities that occur in even the purest material. The electron-defect scattering is approximately temperature independent [16], so we can model its contribution by adding a constant residual resistivity term $\rho_0$ to the total electrical resistivity. We directly use an experimentally extracted residual resistivity value of $\rho_0 = 0.8 \times 10^{-8} \Omega$ [12,18] to represent the typical scale of the effect of these defects. To summarize, for each crystallographic direction, the electrical resistivity $\rho$ in our model is given by:

$$\rho = \left\{ \left[ \left( \frac{\sigma}{\tau} \frac{B}{T^2} \right)^{-1} + CT^2 \sum_{k\lambda} \frac{(\hbar\omega_{k\lambda}/k_B T)^3}{(e^{\hbar\omega_{k\lambda}/k_B T} - 1)(1 - e^{-\hbar\omega_{k\lambda}/k_B T})} \right]^{-1} + \rho_{sat}^{-1} \right\}^{-1} + \rho_0, \tag{17}$$

where $\left( \frac{\sigma}{\tau} \right)$ is obtained from the DFT electronic density of states and associated BTE calculations (electronic DFT-BTE calculations). The values of $B$, $C$, and $\rho_{sat}$ are obtained by fitting to experimental resistivity data.

The total thermal conductivity $\kappa$ is then given by:

$$\kappa = \kappa_{ph} + \frac{\pi^2}{3} \left( \frac{k_B}{e} \right)^2 \frac{T}{\rho}, \tag{18}$$

where $\kappa_{ph}$ and $\rho$ are the results from Eq. (2) and Eq. (17), respectively. In total, we have five fitting parameters in our model for α-U thermal conductivity, which are $B$, $C$, $\rho_0$, $\rho_{sat}$ for the electronic contributions and $A$ for the phonon contributions, and where $C$ and $\rho_{sat}$ are potentially anisotropic, i.e., can have directional dependence.

### 2.1.2 Approximation for U-rich alloys

In dilute alloys, the total resistivity must include the residual resistivity $\rho_{res}$ produced by scattering due to alloy atoms. The total resistivity can be represented using Matthiessen's rule [16]:

$$\rho_{total} = \rho_{pure} + \rho_{res}, \tag{19}$$

where $\rho_{pure}$ is the resistivity of pure α-U using Eq. (17), and $\rho_{res}$ in the binary alloy can be estimated using Nordheim's rule [16]:

$$\rho_{res} = Dc(1-c), \tag{20}$$

where $c$ is the alloying concentration and $D$ is a constant. If we assume the impact of alloying elements on the phonon thermal conductivity is small, the thermal conductivity of dilute U-rich alloys can be calculated using our α-U model of Eq. (18), and the only difference is to add $\rho_{res}$ to the total resistivity formula:

$$\rho = \left\{ \left[ \left( \frac{\sigma}{\tau} \frac{B}{T^2} \right)^{-1} + CT^2 \sum_{k\lambda} \frac{(\hbar\omega_{k\lambda}/k_BT)^3}{(e^{\hbar\omega_{k\lambda}/k_BT}-1)(1-e^{-\hbar\omega_{k\lambda}/k_BT})} \right]^{-1} + \rho_{sat}^{-1} \right\}^{-1} + \rho_0 + Dc(1-c). \tag{21}$$

We apply this model as an approximation for binary U-rich U-Zr and U-Mo alloys in which the U at% is > 78%, and fit the parameter $D$ using Eq. (18) with Eq. (21) for U alloys with different concentrations. This approach is supported by the fact that our calculated results fit experimental thermal conductivity data of U-Zr and U-Mo alloys well at 300-933K, which we show in Section 3.4.

## 2.2 Structural characteristics of α-U

### 2.2.1 Anisotropy

Bulk α-U crystallizes in an orthorhombic structure (Space group: *Cmcm*, No. 63) [30]. Due to its anisotropic properties, we have evaluated our computational model for three different crystallographic directions: [100], [010] and [001]. For each crystallographic direction, we used Eq. (17) to fit experimental data of single crystal α-U resistivities, then calculated the anisotropic thermal conductivity with Eq. (18). To make comparisons with polycrystalline experimental data, we use the fact that the upper bound (UB) and lower bound (LB) of the thermal conductivity and resistivity are given by [31]:

$$\kappa_{UB} = \frac{1}{3}(\kappa_{100} + \kappa_{010} + \kappa_{001}), \tag{22}$$

$$\kappa_{LB} = 3\left(\frac{1}{\kappa_{100}} + \frac{1}{\kappa_{010}} + \frac{1}{\kappa_{001}}\right)^{-1}, \tag{23}$$

$$\rho_{UB} = \frac{1}{3}(\rho_{100} + \rho_{010} + \rho_{001}), \tag{24}$$

$$\rho_{LB} = 3\left(\frac{1}{\rho_{100}} + \frac{1}{\rho_{010}} + \frac{1}{\rho_{001}}\right)^{-1}. \tag{25}$$

We use the simple average of the upper and lower bounds to estimate the total thermal conductivity $\kappa_{total}$ and total resistivity $\rho_{total}$:

$$\kappa_{total} = \frac{1}{2}(\kappa_{UB} + \kappa_{LB}), \tag{26}$$

$$\rho_{total} = \frac{1}{2}(\rho_{UB} + \rho_{LB}), \tag{27}$$

We note here that the differences between the upper bound and lower bound for both thermal conductivity and resistivity are small for our U model: 1% difference for T>450K, and 1%-3% difference for 43K<T<450K.

### 2.2.2   Phase stability

We note here there are other U phases besides α-U that are stable at temperatures below 43 K and above 933 K. A series of three low-temperature charge density wave (CDW) structural phase changes occur below 43K [32,33], and the phase transition to β-U occurs at 933K [33]. Therefore, the valid temperature range of our model is between 43 K to 933 K, which is the temperature range where α-U is stable. The typical operating temperature of a fast nuclear reactor is about 600 K, which is well within the temperature range our model can accurately capture. For the remainder of this work, all models are evaluated and all results are plotted using this relevant temperature range of 43 K to 933 K.

### 2.2.3   Perfect crystal approximation

In this study, our DFT calculations model α-U as an ideal crystalline material. Therefore, our phonon DFT calculations of α-U do not directly include effects on the thermal conductivity due to point and extended (dislocation and grain boundary) defects, and the effects of these defects (electron-defect, phonon-grain boundary and phonon-defect scattering) are counted in our model separately, as mentioned in Section 2.1.1. The magnitude of the defect effect depends on the defect concentration present in the experimental samples. However, since the electrical resistivity in our model is fit to single-crystal resistivity data [34], [35], the effects of electron-grain boundary scattering are not included in our model.

With respect to the effect of electron-grain boundary scattering, we will also show in Section 4.2 by comparing the single-crystalline and polycrystalline resistivity data, that both types of samples display the same temperature dependence with a small difference (~5%) in resistivity values, at least for a typical grain size that is ~0.015mm in diameter [18]. These results indicate that structural differences between single crystal and polycrystalline samples produce only a small difference in the resistivity. Thus, the effect of electron-grain boundary scattering on the thermal conductivity is expected to be small at high temperature. Our model can therefore be considered accurate within 5-10% near 600K for materials with defects concentrations and grain sizes similar to those used in our fitting and the above discussion describes how each term might be affected if significantly different defect levels are considered.

## 2.3   DFT calculations of α-U

All DFT calculations were performed with periodic boundary conditions with a plane-wave basis set using VASP. Initial atomic coordinates for α-U were obtained from the orthorhombic structure (Space group: *Cmcm*, No. 63) in Ref. [30,37]. The electron-ion interaction of U uses the projector augmented wave (PAW) method [38] as implemented by Kresse and Joubert [39]. The valence electron configuration for the U pseudopotential was $6s^2 6p^6 7s^2 5f^3 6d^1$. The exchange correlation functional parameterized in the generalized gradient approximation (GGA) [40] by Perdew, Burke, and Ernzerhof (PBE) [41] was used. The plane wave cutoff energy was set to 450 eV, and the stopping criteria for self-consistent loops were 1 meV per cell for electronic and ionic relaxation. The lattice constants from our structure relaxation calculations agree well with Beeler et al.'s calculation results [42](difference < 0.3%) and Barrett et al.'s experimental data [37], which values are presented in Appendix I. For phonon calculations, the phonon band structure was simulated with a 4×4×4 supercell (128 atoms) and a 4×4×2 Monkhorst-Pack [43] k-point grid. Anharmonic force constant calculations were performed using the small displacement method [44] with a 2×2×2 supercell (16 atoms) and a 10×10×10 Monkhorst-Pack k-point grid. The phonon contribution to the thermal conductivity (phonon-phonon scattering) was obtained by calculating the phonon BTE using Phono3py [6] with a 13×13×13 q-point mesh, resulting in a convergence error of phonon thermal conductivity of < 10%. For electron calculations, band structure calculations were performed with a 2×2×2 supercell (16 atoms) and a 23×23×12 Monkhorst-Pack k-point grid. The electron-electron scattering portion of the electronic contribution to the thermal conductivity was obtained using the BoltzTraP [23] software to conduct BTE calculations in the RTA with a convergence error of < 5%.

## 3    Results

### 3.1    Phonon contribution to U thermal conductivity

The phonon DFT calculation results for α-U are plotted in Figure 1 and Figure 2. Figure 1 illustrates the calculated α-U phonon dispersion curves along [100], [010] and [001] directions at equilibrium calculated by VASP and Phono3py using the small displacement method. The dots in the figure show Crummett et al.'s experimental data for inelastic neutron scattering at room temperature [45]. Our calculation results agree well with Bouchet's calculation [46] and Yang et al.'s calculation [47]. In addition, Bouchet and Yang et al.'s work suggested that the discrepancies between calculations and experiments of the optical branches of [010] and [001] directions are due to the large uncertainties of the experimental measurements for these modes and the temperature difference between the DFT calculations (which are at zero temperature) and experiments conducted at room temperature.

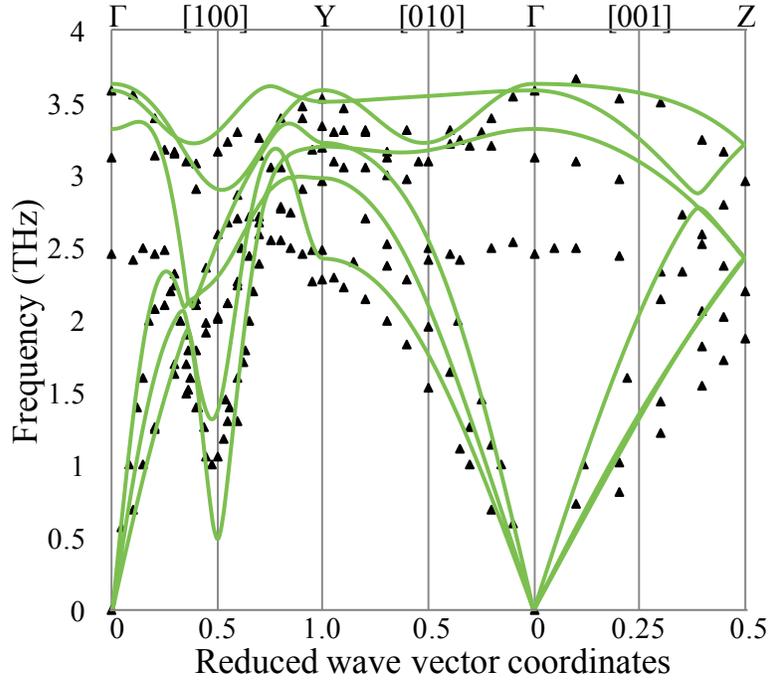

**Figure 1.** The calculated phonon dispersion curves along [100], [010] and [001] directions for α-U. The dots represent Crummett et al.'s experimental data [49] obtained from inelastic neutron scattering.

Figure 2 illustrates the phonon-phonon scattering contribution for α-U in the temperature range of 43K to 933K calculated using the phonon dispersion data of Figure 1 and the phonon BTE in Eq. (2), where for the phonon relaxation time $\tau_{ph}$ in Eq. (3) only the phonon-phonon scattering contribution $\tau_{ph-ph}$ is included. The curves with different colors correspond to the phonon contribution to the thermal conductivity along different crystallographic directions in the anisotropic α-U structure. For all crystallographic directions, the calculated phonon portion of the thermal conductivity decreases with increasing temperature.

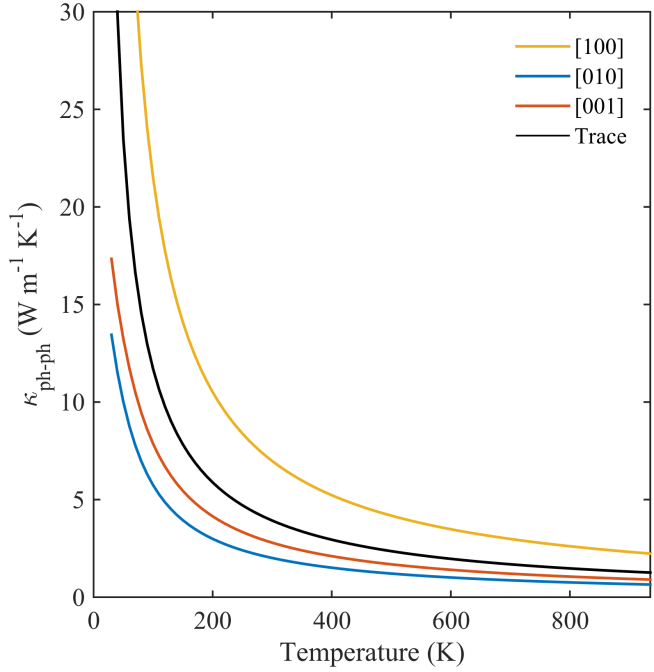

**Figure 2.** Plot of the phonon contribution to the α-U thermal conductivity in the temperature range of 43K to 933K. The different curves correspond to the phonon contribution to the thermal conductivity along different crystallographic directions and the black curve indicates the average over the different directions. Only the phonon-phonon scattering is included in these values.

The phonon-grain boundary and phonon-defect scattering contributions are added into $\tau_{ph}$ using Eq. (3), Eq. (4) and Eq. (5). As discussed in Section 2.1, the value of grain size used is $L$=0.015mm. The parameter $A$ is fit to the experimental phonon thermal conductivity data at 54K: $\kappa$=32.3 W/m-K from Hall et al. [12], $\rho = 0.60\times10^{-7}\Omega$m in our model (shown in Section 3.2), therefore $\kappa_e$=22.0 W/m-K from Eq. (6), and finally $\kappa_{ph}$=10.3 W/m-K at 54K from Eq. (1). Using this data point, $A$ is fit to be $A = 1.0\times10^{-2}(ps)^3$, and the calculated phonon scattering contributions for polycrystalline α-U are shown in Figure 3. The different curves correspond to the different phonon scattering contributions to the thermal conductivity and the black curve indicates the phonon thermal conductivity obtained from Eq. (2). At the fast nuclear reactor working temperature near 600K, the phonon thermal conductivity is dominated by phonon-phonon scattering, and is <4.5% of total thermal conductivity. Therefore, the influence of phonons on the total thermal conductivity can be considered negligible for most applications. A detailed discussion about the contribution of different phonon scattering mechanisms is presented in Section 4.1.

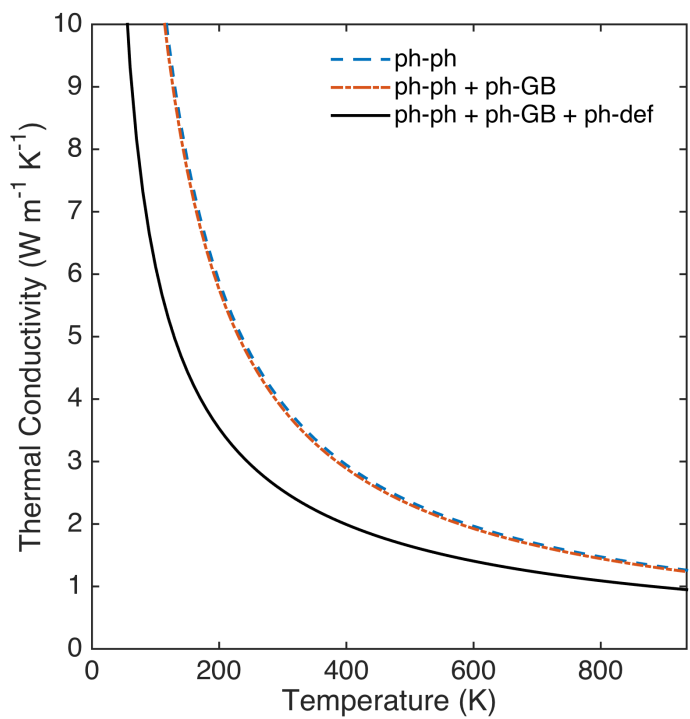

**Figure 3.** Plot of the phonon scattering contributions to the α-U thermal conductivity in the temperature range of 43K to 933K. The different curves correspond to the different phonon scattering contributions to the thermal conductivity and the black curve indicates the phonon thermal conductivity obtained from Eq. (2).

### 3.2 Electronic contribution to U thermal conductivity

Figure 4 contains a plot of $(\frac{\sigma}{\tau})$ (electrical conductivity divided by relaxation time) of electrons along different crystallographic directions in α-U over the temperature range of 43K to 933K, as determined from the electronic BTE of Eq. (9). Overall, the value of $(\frac{\sigma}{\tau})$ shows a very weak temperature dependence of < 10% change over the entire range of 43K to 933K.

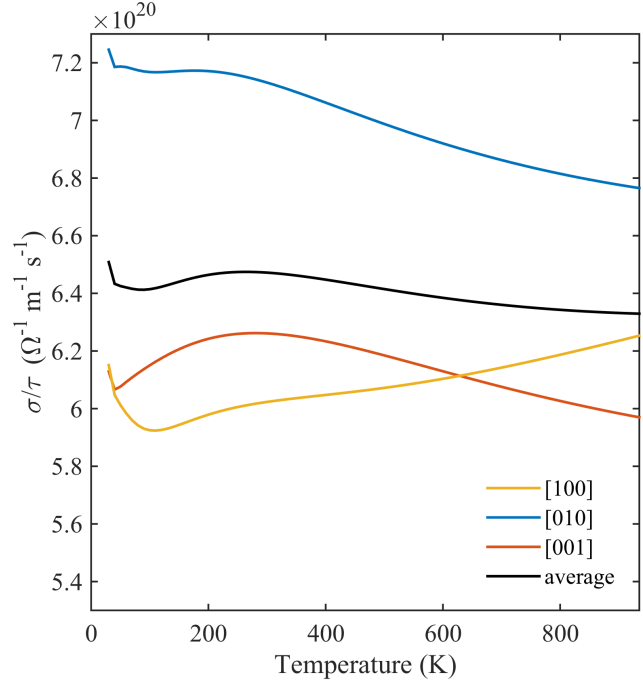

**Figure 4.** Plot of $(\frac{\sigma}{\tau})$ (electrical conductivity divided by relaxation time) of electrons in α-U over the temperature range of 43K to 933K. The curves with different colors indicate the value of $(\frac{\sigma}{\tau})$ along different crystallographic directions and the black curve indicates the average over the different directions.

Now that $(\frac{\sigma}{\tau})$ has been determined, we can fit *B*, *C*, $\rho_{sat}$ and $\rho_0$ to model single crystal electrical resistivity data using the complete resistivity model in Eq. (17). Fitting parameters are calculated by minimizing the root-mean-square error (RMSE) between the model and all experimental data for single crystal resistivity, and the standard deviations of the fitting parameters are obtained from the coefficient covariance matrix from fitting [48]. All fits are performed with the "nlinfit" subroutine in MATLAB (MATLAB and Statistics Toolbox Release 2015a, The MathWorks, Inc., Natick, Massachusetts, United States.). The fitted parameter values and their standard deviations are given in Table 1. Figure 5 contains single crystal α-U experimental resistivity data along different crystallographic directions together with our full resistivity model of Eq. (17) fitted to these data. From the standard deviations of all fitting parameters the standard deviations of the predicted single crystal resistivity are calculated via a propagation of error formula [49], providing an estimate of the uncertainty of the model predictions. One standard deviation for all model predictions is shown in Figure 5 as shaded areas, from which the model uncertainties are all within 10% of the calculated single crystal resistivity. However, this uncertainty is large enough that almost all experimental data are within the uncertainty of our model.

**Table 1.** Fitting parameters obtained from fitting Eq. (17) to anisotropic single crystal α-U experimental resistivity data.

|  | [100] | [010] | [001] |
|---|---|---|---|
| $C(10^{-9}\Omega\text{mK})$ | 11.3±0.4 | 6.3±0.2 | 8.8±0.2 |
| $\rho_{sat}(10^{-7}\Omega\text{m})$ | 7.3±0.4 | 8.5±0.9 | 9.7±0.7 |
| $B(10^7\text{s}\cdot\text{K}^2)$ | 0.3±0.2 | | |

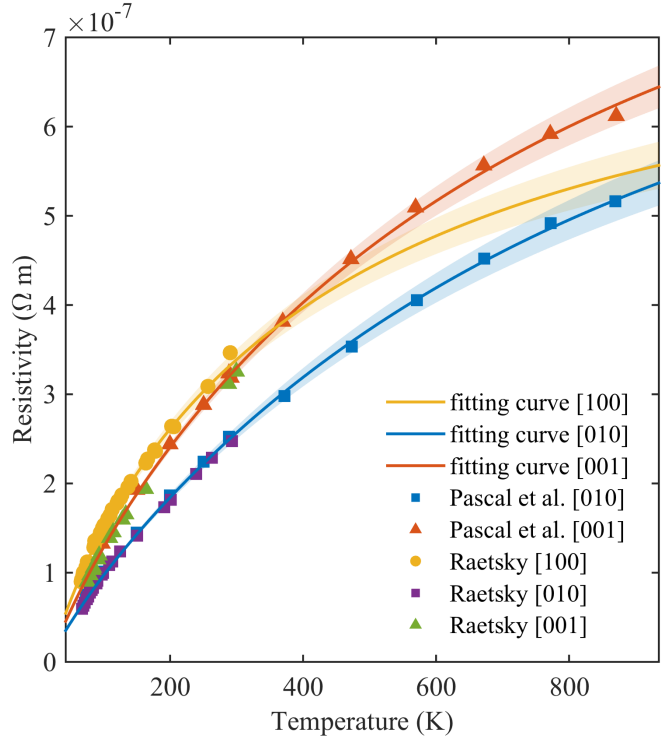

**Figure 5.** Plot of fitted resistivity model (using Eq. (17), solid curves) of α-U resistivity from 43K to 933K, compared with single crystal α-U experimental resistivity data (symbols). The curves with different colors represent the fitted resistivity along different crystallographic

directions, and the shaded areas represent the error ranges. The experimental data for the [010] and [001] directions were obtained from Pascal, et al. [36] and Raetsky [35], and the [100] data was from Raetsky [35].

We use Eq. (27) to estimate the electrical resistivity of polycrystalline α-U. The calculation results of resistivity are shown in Figure 6, together with a comparison to experimental data. The filled circles represent experimental data of single crystal α-U, and the open circles represent experimental data of polycrystalline α-U. The solid black curve ($\rho_{total}$) is the calculated resistivity of α-U, and the dashed orange ($\rho_{e\text{-}ph}$), purple ($\rho_{e\text{-}e}$) and blue ($\rho_0$) curves are the electron-phonon, electron-electron, electron-defect scattering contributions, respectively (the equations used to estimate $\rho_{e\text{-}ph}$, $\rho_{e\text{-}e}$ and $\rho_0$ are shown in Appendix II). The shade area is the error range of calculated resistivity. The polycrystalline and single crystal α-U resistivity data have the same temperature dependence. The average of the single crystal data over different directions falls within the spread of the polycrystalline data. This result suggests grain boundary effects are smaller than the uncertainty introduced by different experiments and by our averaging over the single crystal data, which are discussed later in Section 4.2. Here and in some later sections of the paper we will calculate errors between model and experimental data where the experimental data is often clustered in certain temperature ranges. In such comparisons, when possible we extrapolate the experiments through linear interpolation onto an approximately uniform temperature grid so that errors represent uniform sampling over temperature. These temperature grids can be found for each case in the Appendices. The mean error (ME) and root-mean-square error (RMSE) between our model and average polycrystalline data are ($0.02 \pm 0.04$)$\times 10^{-7}\Omega$m and $0.15 \times 10^{-7}\Omega$m, respectively, which are <5% of the total resistivity value. All values of resistivity from our calculation results (Appendix III, Table A2) and the known experimental measurements (Appendix IV, Table A4) are provided in the Appendices.

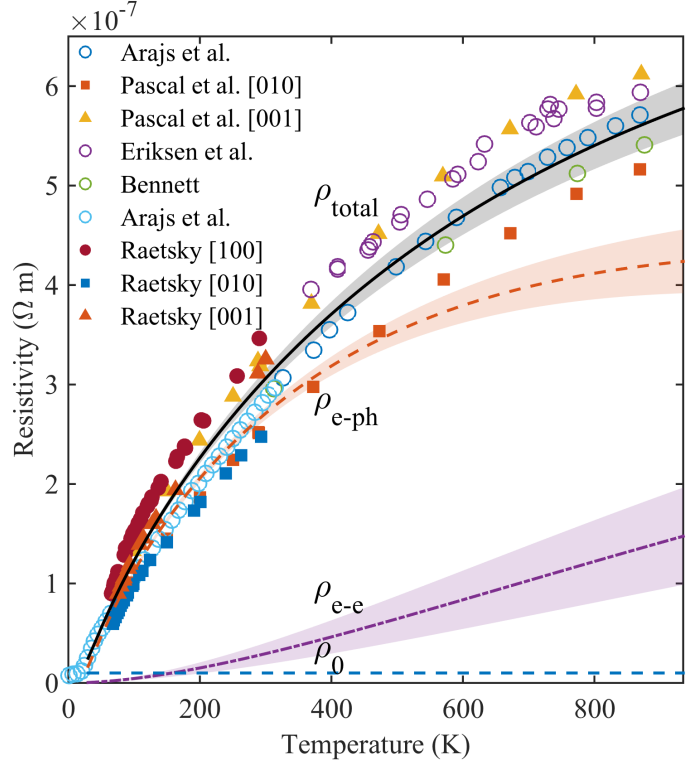

**Figure 6.** Calculation results of α-U resistivity from 43K to 933K (curves), compared with experimental data (symbols). The filled circles represent experimental data of single crystal α-U from Ref. [35,36], and the open circles represent experimental data of polycrystalline α-U from Refs. [18,50–52]. The solid black curve ($\rho_{total}$) is the calculated resistivity of α-U, and the dashed orange, purple and blue curves, labeled $\rho_{e-ph}$, $\rho_{e-e}$ and $\rho_o$, respectively, are the electron-phonon, electron-electron scattering contributions and residual resistivity due to electron scattering with point defects.

### 3.3   Total model for U thermal conductivity

Using Eq. (18), Eq. (26), our calculated phonon thermal conductivity and fitted resistivity results from Table 1, the thermal conductivity of α-U was calculated. The values are tabulated in Table A3 of Appendix III, along with all available experimental values from the literature in Table A5 of Appendix IV. Figure 7 shows the anisotropic thermal conductivity results of α-U compared with experimental polycrystalline thermal conductivity data. To our knowledge, no thermal conductivity data of single crystal α-U exists in the literature. Almost all of the experimental polycrystalline data points are within our predicted α-U thermal conductivity curves. In Figure 7 and Figure 8, the solid black curve ($\kappa_{total}$) is the thermal conductivity of polycrystalline α-U, estimated using an average of different crystallographic directions (Eq. (26)). In Figure 8, the dashed orange ($\kappa_{ph}$) and purple ($\kappa_e$) curves illustrate the phonon contribution and electronic contribution, respectively. For the entire temperature range, the ME and RMSE between our model and experimental thermal conductivity data are 0.09±0.11 W/m-K and 0.41 W/m-K, respectively. Our model is within the range of reported experimental data and shows good overall agreement.

Despite the good overall agreement of our model with the experimental data, we obtain slightly higher thermal conductivity values than some experimental data for 300-700K, slightly lower thermal conductivity values for 750-933K (see errors in Table A3 of Appendix III), and show some dramatic errors below about 300K for Tyler et al.'s data [10] and below about 100K for Hall et al.'s data [12]. These issues are discussed respectively in Section 4.3.

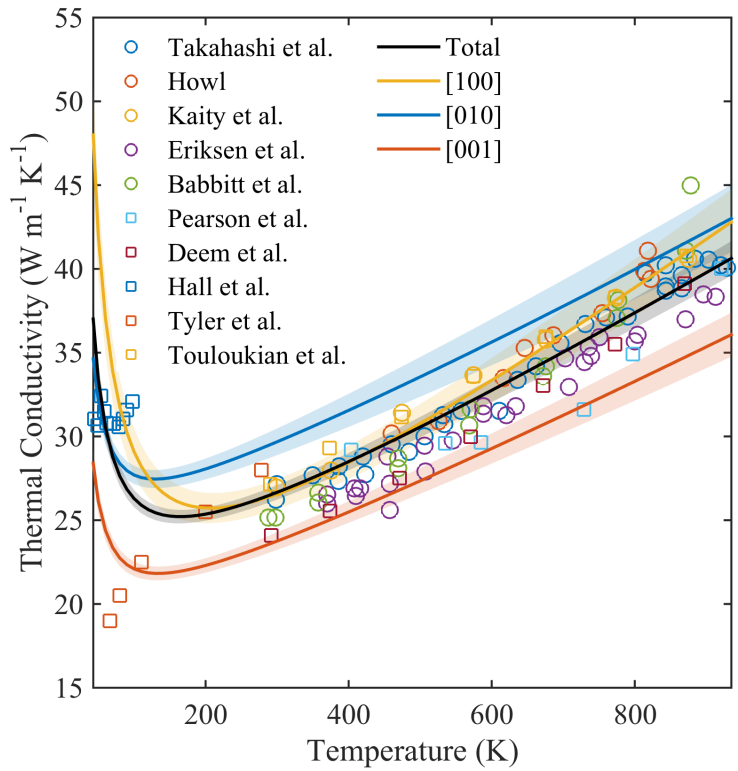

**Figure 7.** Plot of our thermal conductivity model from 43K to 933K (solid curves), compared with polycrystalline α-U experimental thermal conductivity data (symbols). The curves with different colors represent the calculated thermal conductivity along different crystallographic directions. The experimental thermal conductivity data were obtained from the following references: [10–13,50,53,54].

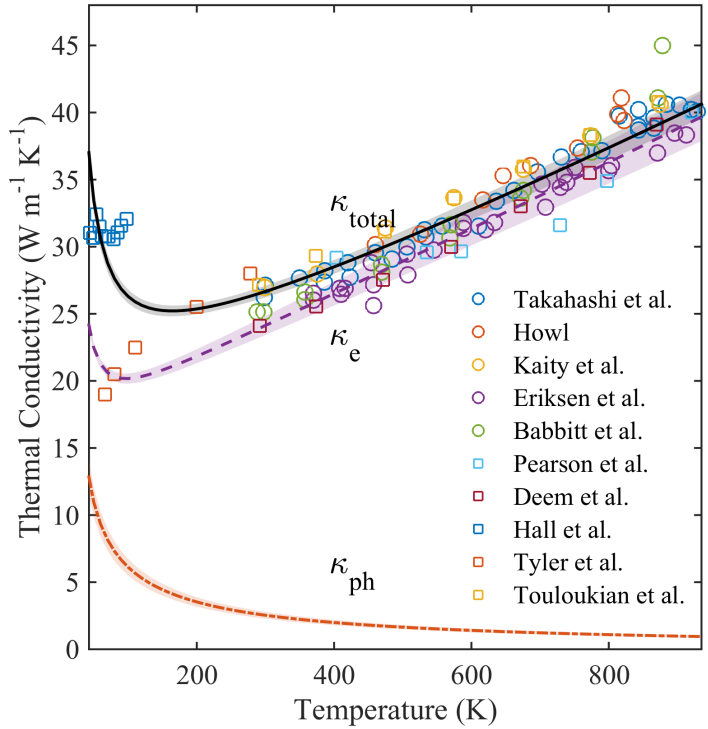

**Figure 8.** Calculation results of α-U thermal conductivity from 43K to 933K (curves), compared with polycrystalline α-U experimental thermal conductivity data (symbols). The solid black curve ($\kappa_{total}$) is the calculated thermal conductivity of α-U, and the dashed orange and purple curves, labeled $\kappa_{ph}$ and $\kappa_e$, respectively, are the phonon contribution and electronic contribution to the thermal conductivity, respectively. The experimental thermal conductivity data were obtained from the following references: [10–13,50,53,54].

### 3.4 Extended thermal conductivity model for binary U-rich alloys

The thermal conductivities of binary U-rich U-Zr and U-Mo alloys from 300K to 933K were calculated by fitting the full thermal conductivity equation (Eq. (18)) with the alloy resistivity equation (Eq. (21)) to experimental data. The thermal conductivity values are tabulated in Table A6 and A7 of Appendix V, and experimental data of thermal conductivity for alloys is tabulated in Table A8 and A9 of Appendix VI. Figure 9 and Figure 10 contain U-Zr and U-Mo experimental thermal conductivity data with our alloy thermal conductivity model fitted to these data, respectively. The fitting parameters are the same as values in Table 1 except for the addition of the parameter $D$ in Eq. (21). The values of $D$ obtained from the fit are: $D_{UZr} = (0.97 \pm 0.08) \times 10^{-6}\Omega$m, $D_{UMo} = (1.51 \pm 0.13) \times 10^{-6}\Omega$m. The overall RMSE values in our fits are 1.3 W/m-K and 1.6 W/m-K for U-Zr and U-Mo, respectively. The overall ME values in our fits are $(0.15 \pm 0.18)$W/m-K and $(0.29 \pm 0.24)$ W/m-K for U-Zr and U-Mo, respectively. We note here that in theory this model only applies for U-rich alloys with a dilute amount of impurities. However, the agreement between our fitted model results and experimental data shows that despite the alloy composition being somewhat out of the dilute regime (we consider U at% > 78%) this model is still a good approximation for U-Zr and U-Mo alloys. We also note that the typical alloy components in actual fuels are U-22at%Zr in the Experimental Breeder Reactor-II (EBR-II) in Idaho, USA [8], and U-15.7at%Mo in the Belgian Reactor 2 (BR2) in Mol, Belgium [55]. These Zr and Mo composition ranges are covered by our model. Based on experimental data, Kim et al. developed empirical formulas for the thermal conductivity of the U–Zr [3,56] and U-Mo alloys [4,57] applicable to any composition, which are shown as dashed curves in Figure 9 and Figure 10, respectively. Comparing to the same data, the overall RMSE values for Kim et al.'s empirical formulas are 1.6 W/m-K for both U-Zr and U-Mo. Thus, based only on the RMSE values, for U-Zr our model is slightly better than Kim et al.'s formula, and for U-Mo our model is as good as Kim et al.'s formula. However, while both our and Kim et al.'s models are based on the thermal conductivity of pure α-U, Kim et al.'s formulas use five parameters to fit the thermal conductivity of alloys, whereas we only use one. This suggests that our model may give a more complete physics-based representation of the contributions to the thermal conductivity, and that our modeling approach can be extended to complex materials like U alloys.

One advantage of our model having fewer fitting parameters relative to a typical empirical model, e.g., that for U-Zr from Kim, et al., is that less experimental data is needed to construct our model than might be needed for an empirical model. Here, we demonstrate the extensibility of our model beyond $\alpha$-U to alloy systems with very limited data by showing we can achieve good agreement with experimental U-Zr thermal conductivities in the limit where only a single experimental data point is available. Our model can capture experimental thermal conductivity data of the U-Zr alloy system using a single data point because there is only one fitting parameter in our alloy model. Using each of the U-Zr experimental data points as fitting data, we predict the thermal conductivity of U-Zr and calculate the RMSE relative to the experimental data. One example is given in Figure A1 in Appendix VII, and the mean RMSE value from fitting to each single U-Zr experimental data point and predicting the others is 2.1 W/m-K. This quite small error result shows that even with one data point, our model can still accurately predict the thermal conductivity for alloys. Thus, for the case where little experimental data is available, our model can still predict reliable results compared to experiment, a result that would not be possible with a more empirical model like that used by Kim, et al. In general, obtaining high-quality experimental thermal conductivity measurements for alloys, especially U alloys, is expensive and difficult. Thus, it is potentially very useful to know the thermal conductivity value within 10% error for a wide temperature and concentration range via one or a few experimental data points, as we have demonstrated can be done with our model for U-Zr.

Overall, the current alloying model developed in this work is still somewhat approximate. More specifically, we don't include the impact of alloying elements on the phonon thermal conductivity and use a one-parameter fitting formula for residual resistivity with only up to quadratic concentration dependence and with no temperature dependence. Adding the effects of alloying on the phonon thermal conductivity and incorporating a more complex alloying resistivity formula, e.g., with temperature dependence and/or higher order terms in concentration, could produce a more accurate model compared to experimental data for U alloys, including U-Zr and U-Mo. Developing such a model will be the subject of future work.

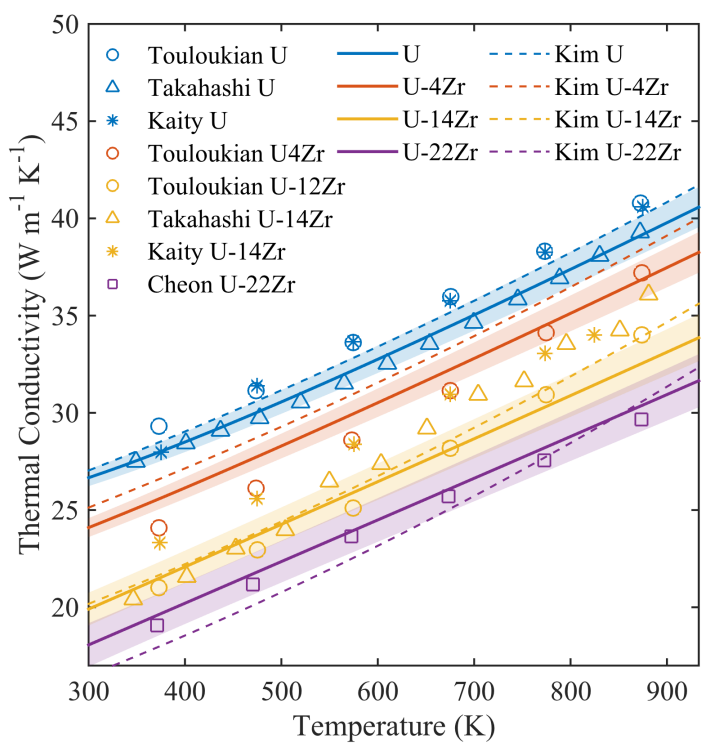

**Figure 9.** Calculation results of U-rich U-Zr alloys thermal conductivity from 300K to 933K (solid curves), compared with U-Zr experimental thermal conductivity data (symbols). The dashed lines are values based on empirical models developed by Kim, et al. [3]. The curves and symbols with different colors represent the U-Zr thermal conductivities in different Zr at%. The experimental thermal conductivity data were obtained from the following references: [10,11,13,56,58].

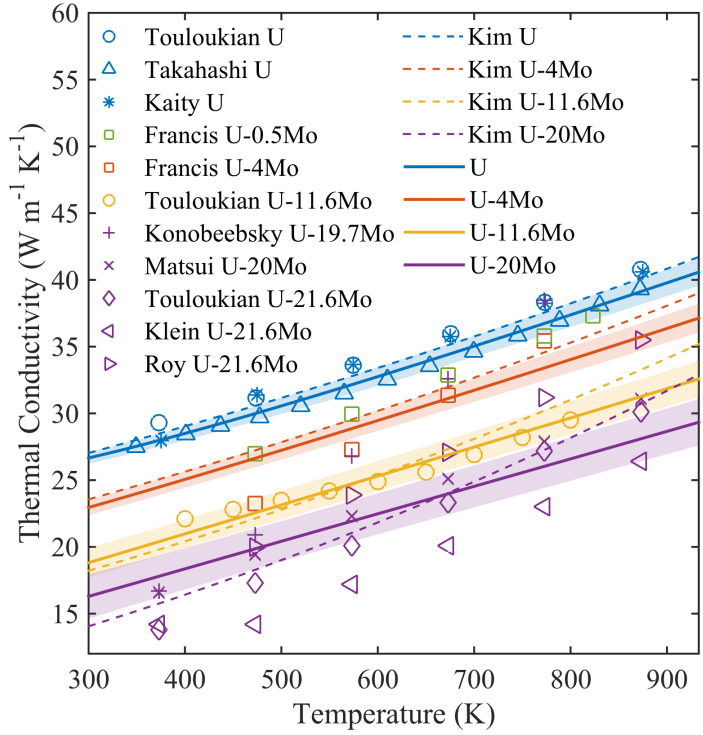

**Figure 10.** Calculation results of U-rich U-Mo alloys thermal conductivity from 300K to 933K (solid curves), compared with U-Mo experimental thermal conductivity data (symbols). The dashed lines are values based on empirical models developed by Kim, et al. [4]. The curves and symbols with different colors represent the U-Mo thermal conductivities in different Mo at%. The experimental thermal conductivity data were obtained from the following references: [10,11,13,57,59–63].

## 4    Discussion

### 4.1    Phonon thermal conductivity

Our model for phonon thermal conductivity contains phonon-phonon, phonon-grain boundary, and phonon-defect scattering, but other scattering processes for phonons are neglected, specifically phonon-electron, phonon-dislocation, and phonon-isotope scattering. The contributions from these mechanisms can be easily included in our formalism by adding their respective phonon relaxation time terms in our model through Eq. (3). The phonon-electron and phonon-dislocation scattering are expected to only limit the phonon thermal conductivity at low temperature and are negligible above room temperature [16]. Recent ab-initio calculations for multiple metals like Al and Ag support the assertion that the phonon-electron scattering is negligible above room temperature [64]. Ab-initio studies also support the assertion that the phonon-dislocation scattering is negligible above room temperature for dislocation densities < $10^9$ cm$^{-2}$ [65], and the dislocation density for annealed metals is ~$10^7$-$10^8$ cm$^{-2}$ [66]. Therefore, these contributions are appropriate to exclude for the conditions of focus in this work. For phonon-isotope scattering, we estimate its contribution by adding the phonon-isotopic scattering relaxation time following Togo et al.'s approach [6] with the mass variance parameter of U from Laeter et al.'s values [67], and the results are discussed in the next paragraph.

Above 600K, the phonon thermal conductivity is below 1.4 W/m-K and the experimentally measured thermal conductivity of α-U is above 30 W/m-K (Figure 3). Therefore, the phonon contribution to the total thermal conductivity at reactor operating conditions in relatively pure α-U is likely negligible for most considerations, although this could change in irradiated systems and fuel alloys. Therefore, it is still of importance to evaluate which scattering mechanism dominates the phonon thermal conductivity. Although some parameters in our model are only accurate for specific samples in specific experiments, e.g. the grain size $L$ and the constant $A$ (which depends on defect concentration), the model represents typical parameter values and can be easily adapted to new parameter values as needed. To separate the values of different phonon scattering contributions, the Matthiessen approximation [16] is applied:

$$\frac{1}{\kappa^{ph}} = W^{ph-ph} + W^{ph-gb} + W^{ph-def} + W^{ph-iso} \qquad (28)$$

where $\kappa^{ph}$ is the phonon thermal conductivity, and $W^{ph-ph}$, $W^{ph-gb}$, $W^{ph-def}$ and $W^{ph-iso}$ are the estimated thermal resistivity values of phonon-phonon, phonon-grain boundary, phonon-defect and phonon-isotope scattering, respectively. These thermal resistivity values can be obtained from Eq. (2) with different phonon scattering contributions. From Eq. (28) it is clear that $\kappa^{ph}$ is controlled by the scattering with the largest value of thermal resistivity. Examination of the results in Figure 3 shows that phonon-defect scattering plays a major role in controlling the phonon thermal conductivity when $T$<70K, while for $T$>70K, phonon-phonon scattering plays a major role. Near the reactor operating temperature of 600K, the phonon-phonon scattering controls ~71% of the total phonon thermal conductivity (i.e., $W^{ph-ph}/\left(W^{ph-ph} + W^{ph-gb} + W^{ph-def} + W^{ph-iso}\right) \approx 71\%$), and the phonon-defect, phonon-grain boundary and phonon-isotope scattering controls ~27%, 2% and 0.05% of total phonon thermal conductivity, respectively.

## 4.2   Resistivity and electrical thermal conductivity

As mentioned in Section 2.1.1, the electron-phonon scattering contribution $\rho_{e-ph}$ uses semi-empirical fitting model instead of DFT calculations. The DFT approaches for realistic systems generally follow those taken in, e.g., Savrasov et al.'s [5] and Verstraete's [68] calculations. In these approaches, effects of anharmonicity effects are not taken into account at high temperature, so the temperature range is limited by $T < 2\theta_{tr}$, where $\theta_{tr}$ is the average phonon frequency [5,69]. For α-U, $\theta_{tr} \approx 130K$ from our phonon calculations, thus these DFT-based approaches to calculate $\rho_{e-ph}$ do not cover the reactor working temperature near 600K. Although some approaches can calculate the phonon anharmonicity, like Phono3py [6] and phonon Monte Carlo [70], a robust approach that includes phonon anharmonicity into the calculation of $\rho_{e-ph}$ is not presently available and validated for complex systems like U.

Therefore, we did not pursue DFT calculations of $\rho_{e-ph}$ and instead fit a semi-empirical model to the available experimental data and phonon spectrum calculation results with Eq. (15).

To check the validity of Eq. (15), we compare the Debye temperature of resistivity and heat capacity, as a descriptor, from our calculation results and experimental data, and they are discussed below. From our calculation results, Eq. (15) is in close agreement with the Bloch-Gruneisen formula (Eq. (14)) with $\theta_D^R \approx 115K$ for α-U. $\theta_D^R$ values have been estimated from resistivity measurements by Lee, et al. and by Meaden as 121K [71] and 170K [18], respectively. We show excellent agreement with Lee, et al., but are somewhat lower than Meaden. The disagreement between our and Meaden's estimated values is expected to come from two sources. First, our Eq. (15) is not exact, which will introduce errors when fitting Eq. (14) and $\theta_D^R$ to the results of Eq. (15) when compared to fitting to presumably more exact experimental data. However, if approximations in Eq. (15) were the dominant source of error they should have led to disagreements with both Meaden and Lee, et al. Second, the experimental resistivity data contains contributions not included in our single crystal model, specifically, the small resistivity step due to the phase transition near 43K [18], the lattice strains induced by heating and cooling processes near the phase transition [72], and the impurities that can influence the temperature dependence of resistivity at low temperature [18]. These contributions may be more significant in Meaden's experiments that Lee, et al., driving the larger differences. While the difference between the $\theta_D^R$ fit in our model and Meaden's value appears large, the resistivity of the Bloch-Gruneisen formula is approximately independent of $\theta_D^R$ when $T > 0.4\theta_D^R$, so a discrepancy in $\theta_D^R$ only affects the resistivity values at low temperature and does not significantly influence our model parameterization, or model predictions above about 100K. The $\theta_D^R$ obtained from resistivity measurements, both in our model and the values estimated from experiments, are lower than the Debye temperature obtained from heat capacity ($\theta_D^C$), which is ~200K [30] from experimental data and is 176K calculated from our phonon spectrum. This difference is not unexpected because the phonon bands of α-U are not a Debye phonon spectrum, and the $\theta_D^R$ and $\theta_D^C$ values are affected differently by the Debye approximation.

To estimate the magnitude of the effect of grain boundaries on resistivity, we should calculate the RMSE between single- and polycrystalline resistivity data. However, due to the lack of resistivity along [100] direction at 300-933K, we can't directly calculate the average of single crystal resistivity data. As our resistivity model is fitted to single crystal data, we use the RMSE value between our model and average polycrystalline data to estimate the RMSE between single- and polycrystalline resistivity data, which is <5% as mentioned in Section 3.2. Such a small difference of the resistivity between single crystal and polycrystalline samples suggests that the effects of grain boundaries are small and our estimation of averaging the resistivity (and electronic contributions to thermal conductivity) along different crystallographic directions as a proxy for a polycrystalline resistivity (and electronic contributions to thermal conductivity) value is reasonable.

From the results of our fitted resistivity model, we can clearly see which physical contribution dominates the resistivity of α-U. In Figure 6, the electron-phonon scattering contribution of resistivity is larger than both the electron-electron scattering part and residual resistivity, and dominates the temperature dependence of resistivity from 43 to 933K.

## 4.3   Total thermal conductivity

As mentioned in Section 3.3, we have slightly higher thermal conductivity values than some experimental data for 300-700K, slightly lower thermal conductivity values for 750-933K (see errors in Table A3 of Appendix III), and show some dramatic errors below about 300K for Tyler et al.'s data [10] and below about 100K for Hall et al.'s data [12].

To explain the discrepancy with Tyler et al.'s data we note that at low temperatures, defects (point defects, impurities and dislocations) can significantly decrease the thermal conductivity of α-U due to their influence on both the phonon and electronic component of the thermal conductivity. Our calculation results show that phonon-defect scattering dominates the phonon thermal conductivity at low temperature. It also has been shown experimentally by Hall et al.'s experiments [12] that the α-U thermal conductivity values increase more than 40% at low temperature after annealing. The fact that annealing tends to remove defects from a material suggests that the samples used by Tyler et al. contain a sizeable number of defects, which decrease both the phonon and electronic contributions to the thermal conductivity. This hypothesis is also supported by the residual-resistance ratio (RRR) of the Hall and Tyler samples, which is 28 for Hall et al.'s, and 10 for Tyler et al.'s [59]. RRR is the ratio of the resistivity at 300K and at temperature close to 0K, and usually a higher RRR value indicates a sample with lower residual resistivity and fewer defects. Therefore, Tyler et al.'s sample may have a higher concentration of defects. Although the origin of the discrepancy between our model and Hall et al.'s data is not clear, we note that the latter shows complex temperature dependence at low temperature, which is not supported by the basic theory of thermal conductivity of metals that underlies our modeling [16]. Therefore, we exclude both Tyler et al.'s and Hall et al.'s data in our RMSE calculations. Only one data point from Hall et al.'s data is used, $\kappa$=32.3 W/m-K at 54K, which is used to estimate the correct scale of phonon-defect scattering. Although the scale is correct, the defect concentrations among different experiments are still different, and this may lead to the slight discrepancies at 300-933K.

For the slight discrepancies at 300-933K, they may be due to missing physics in the model, or due to the inaccuracy of the model results for the [100] direction, and we discuss both of these possibilities. A potentially significant piece of missing physics in the model is that our phonon band structures are calculated at 0K and ignore the physics of thermal expansion. Thermal expansion effects will decrease the density of materials and therefore decrease phonon thermal conductivity [73]. The influence of thermal conductivity, estimated from the Leibfried-Schlemaan formula for phonon-phonon scattering [74] and the Gruneisen parameter of U [75] (which are discussed in Appendix VIII), is ~13% near 600K. Thus, thermal expansion effects may make our predicted thermal conductivity slightly higher than actual values by ~1% above 600K, but the change is expected to be negligible. DFT or experimental thermal expansion could be added to a future iteration of the model to reduce any errors associated with these effects. Another piece of missing physics in the model is that we assume the validity of lowest order perturbation treatment for the scattering processes, e.g. the $T^2$ dependence of electron-electron scattering relaxation time and the $\omega^{-4}$ dependence of phonon-defect scattering relaxation time. This assumption is valid for low temperature and may result in the discrepancy at high temperature due to the impact of higher order perturbation. Another source of error between the predicted thermal conductivity and the measured values is that the resistivity prediction for the [100] direction may be inaccurate from 300K to 933K, due to less resistivity data being available in this direction. As the electronic contribution is the majority of the thermal conductivity at high temperature, this inaccuracy in the [100] resistivity data could lead to the observed discrepancies in the total thermal conductivity.

Within our framework, the dominant scattering process of total thermal conductivity of α-U can be easily assessed. In Figure 8, the electronic thermal conductivity is larger than the phonon thermal conductivity from 43K to 933K, and dominates the temperature dependence of thermal conductivity. Thus, electrons are the major heat carriers, and electron-phonon scattering is the dominant interaction of thermal conductivity of α-U, including at the expected working temperature of U-based metallic fuels of around 600K. However, the electron-electron scattering contribution and the phonon contribution (mainly controlled by phonon-phonon and phonon-defect scattering) also play a significant role in setting thermal conductivity values near 800K and 100K, respectively. Near 100K the phonon scattering (mainly controlled by phonon-defect scattering) contributes ~25% of the total thermal conductivity. Near 800K, the electron-electron scattering contributes ~10% of the total thermal conductivity.

To validate that the physical contributions of the model are correct one can fit to just part of the data and assess the ability to extrapolate to the rest of the data. Here we perform such a test based on temperature extrapolation. Specifically, we use only use the resistivity data from the low temperature range of 43-300K for all three crystallographic directions as fitting data, with the same fitting processes and error analysis as before. The results are presented in Figure A2 in Appendix IX. By extrapolating to the high temperature range of 300-933K and comparing to the experimental data from 300-933K, the ME and RMSE are both 2.4 W/m-K, which is ~7% of the total thermal conductivity value. Therefore, our model demonstrates correct temperature dependence of the thermal conductivity even when only low temperature thermal conductivity data are used in the fit, which supports that we have a robust representation of the temperature dependent physics and that our approach could be used effectively in systems with data available only over limited temperature ranges.

## 4.4    Sensitivity and error analyses

As our model is fit to experimental electrical and thermal conductivity data over a range of temperatures from multiple groups, our model is fairly insensitive to errors in one given experimental data point. For temperatures above room temperature, the phonon contribution is dominated by phonon-phonon scattering obtained from DFT, and the electrical contribution is obtained from the Wiedemann-Franz Law to the fitted experimental resistivity. Therefore, our thermal conductivity prediction contains almost no contributions for direct fitting to the thermal conductivity data. Thus, the success of our model when comparing to the measured thermal conductivity is quite encouraging, and our model is largely insensitive to the exact values of the measured thermal conductivity over most of the relevant temperature range. Our model is also expected to be quite insensitive to errors in the band structure of electrons, because the electronic contribution is scaled by a relaxation time that is fit to experimental resistivity data. However, the phonon contributions are sensitive to the phonon spectrum and force constants, which come from DFT that can contain errors. But, as the phonon contribution is small at high temperature, this sensitivity only matters at low temperature and is not critical for this work.

The error ranges, showed in Figure 5 through Figure 10, are calculated from an error propagation formula as mentioned in Section 3.2. The propagated errors are from two sources: error from fitting which is represented by the standard deviations of fitting parameters, and DFT calculation errors. As mentioned in Section 3.2, the standard deviations of fitting parameters are obtained from the coefficient covariance matrix (see SI for values). For nonlinear fitting, the coefficient covariance matrix shows the correlation between fitting parameters. For DFT calculations, the only error included is the convergence error, and the error between DFT calculation and experiments is not included. For phonon spectrum, we compared our calculated results with previous experimental and calculation results in Section 3.1, and the good agreement suggests that this source of error should be small. For phonon force constants and electronic band structure calculations, we have discussed that our model is not very sensitive to these values near reactor operating temperature of about 600K, so we expect their contributions to the overall errors to also be small. As the error ranges (one standard deviation) of different resistivity contributions and thermal conductivity contributions are all < 10% of the total resistivity and total thermal conductivity, respectively, these errors do not affect our prediction of the relative contributions of the different mechanisms

## 4.5    Anisotropy of resistivity and thermal conductivity

From the results of our anisotropic single crystal resistivity model, we predicted that the resistivity along [100] direction becomes lower than the resistivity along [001] direction when T>370K, and nearly identical to the resistivity along [010] direction at ~930K, although all experimental data of [100] direction have larger values than the other two directions in the range 43-300K. These results show that the relative value of the [100] resistivity changes dramatically with temperature, going from the largest, to tied, to the smallest value among the three directions. Such large changes in relative values is not unprecedented in materials exhibiting resistivity saturation, e.g., yttrium and $WO_2$ also show significant changes in the relative values of resistivity along different directions with increasing temperatures [76].

Besides temperature, anisotropy also influences the contributions of different scattering mechanisms to the thermal conductivity. From Figure 5, the electrical resistivity along the [010] direction is smaller than along [100] and [001] directions. This directly leads to the result in Figure 7 that the thermal conductivity along [010] direction is much larger than along the [100] and [001] directions from 150-933K. Also, as shown in Figure 2, the phonon-phonon scattering contribution along the [100] direction is much larger than along the other two directions for T<100K. This leads to the result in Figure 7 that for T<100K, although the resistivity along [100] direction is slightly larger than [001] and [010] direction, the thermal conductivity along the [100] direction is still the largest among three directions.

## 5    Conclusions

A computational, physics-based model of α-U thermal conductivity has been constructed which is based on DFT calculations and experimental data, and which incorporates both phonon and electron scattering contributions, including phonon-phonon, phonon-grain boundary, phonon-defect, electron-phonon, electron-electron and electron-defect scatterings. This is the first model of α-U thermal conductivity that incorporates all known experimental resistivity and thermal conductivity data and separates out contributions from different physical mechanisms. This model provides insight into the different physical factors that govern the thermal conductivity of α-U under different temperature ranges. At 43-933K, electrons are the major heat carriers and the conductivity is generally dominated by electron-phonon scattering. Near reactor operating temperatures of 600K, the phonon contribution is ~4.5% of the overall thermal conductivity, and the electron-phonon scattering contribution is ~80% of the total resistivity. Therefore, the electron-phonon scattering controls ~75% of the overall thermal conductivity near 600K.

Overall, this work serves as a first step to understanding the complex behavior of thermal conduction in metallic U alloy nuclear fuels. Now that a model of pure α-U has been established, it can be used to incorporate more physically realistic and complex effects, such as intentional alloying elements, transmutation products, radiation-induced defects and noble gas bubbles. For example, the thermal conductivity of U-rich U-Zr and U-Mo alloys can be calculated simply by adding a residual resistivity term, with results that agree well with the available experimental data. The usage and extension of this thermal conductivity model of α-U to more complex systems is an important step to gaining a deeper understanding of the thermal conduction characteristics of metallic U-alloy nuclear fuels. This understanding will in turn enable the improved design of temperature control in the future construction of nuclear reactors powered by metallic U-alloy fuels. Finally, the general framework for thermal conductivity of α-U, U-Zr and U-Mo alloys developed in this work can be applied to the generation of ab-initio-based and physics-based semi-empirical thermal conductivity models for other metallic systems.

**Acknowledgements**

We gratefully acknowledge financial support from the Department of Energy Nuclear Energy University Program (NEUP) grant 14-6767. Computations in this work benefitted from the use of the Extreme Science and Engineering Discovery Environment (XSEDE), which is supported by National Science Foundation grant number OCI-1053575.

**Appendix I**: The optimized crystal structure at zero pressure, which is compared with previous results from calculations and experiments and shows a good agreement with previous PBE calculations (difference < 0.3% for lattice constants) and experiment (difference < 1.5% for lattice constants).

**Table A1.** The optimized lattice constants, internal parameters and volume per atom for α-U, compared with Beeler et al.'s first principle calculations [42], Taylor's PW91 pseudopotential calculations [77], Söderlind's full-potential (FP) calculations [78], Crocombette et al.'s norm-conserving (NC) pseudopotential calculations [79] and experimental data at 50 K from Barrett [37].

| | This work (PBE) | Beeler (PBE) | Taylor (PW91) | Söderlind (FP) | Crocombette (NC) | Barrett (exp) |
|---|---|---|---|---|---|---|
| A(Å) | 2.794 | 2.793 | 2.800 | 2.845 | 2.809 | 2.836 |
| B(Å) | 5.844 | 5.849 | 5.896 | 5.818 | 5.447 | 5.867 |
| C(Å) | 4.913 | 4.894 | 4.893 | 4.996 | 4.964 | 4.936 |
| Y | 0.098 | 0.098 | 0.097 | 0.103 | – | 0.102 |
| Volume /atom(Å³) | 20.057 | 19.987 | 20.194 | 20.674 | 19.026 | 20.535 |

**Appendix II:** The equations to separate electron-electron and electron-phonon contribution for total resistivity.

The resistivity for each direction is calculated using (as same as Eq. (17)):

$$\rho^X = \left\{ \left[ \rho_{e-e}^{X,id} + \rho_{e-ph}^{X,id} \right]^{-1} + (\rho_{sat}^X)^{-1} \right\}^{-1} + \rho_0, \tag{A1}$$

where $X$ represents a direction (100), (010), or (001), $\rho_{e-e}^{X,id}$ is calculated using Eq. (8) and Eq. (12), and $\rho_{e-ph}^{X,id}$ is calculated using Eq. (15). We approximately separated electron-electron and electron-phonon contribution by weighting the total resistivity along a direction by the fraction of the ideal resistivity due to each contribution, as show here:

$$\rho_{e-e}^X = \left\{ \left[ \rho_{e-e}^{X,id} + \rho_{e-ph}^{X,id} \right]^{-1} + (\rho_{sat}^X)^{-1} \right\}^{-1} \times \frac{\rho_{e-e}^{X,id}}{\rho_{e-e}^{X,id} + \rho_{e-ph}^{X,id}}, \tag{A2}$$

$$\rho_{e-ph}^X = \left\{ \left[ \rho_{e-e}^{X,id} + \rho_{e-ph}^{X,id} \right]^{-1} + (\rho_{sat}^X)^{-1} \right\}^{-1} \times \frac{\rho_{e-ph}^{X,id}}{\rho_{e-e}^{X,id} + \rho_{e-ph}^{X,id}}. \tag{A3}$$

We used Eq. (27) to calculated total resistivity from anisotropic resistivities of each direction and this equation mixes contributions of $\rho_{e-e}^X$ and $\rho_{e-ph}^X$ in ways that do not allow for a rigorous separation of $\rho_{total}$ into electron-electron and electron-phonon components. Therefore, we instead use a simple average to calculate the total contribution of electron-electron and electron-phonon scattering:

$$\rho_{e-e/e-ph} = \frac{1}{3} \times \left( \rho_{e-e/e-ph}^{100} + \rho_{e-e/e-ph}^{010} + \rho_{e-e/e-ph}^{001} \right). \tag{A4}$$

This average means that $\rho_{total}$ is not rigorously equal to $\rho_{e-e} + \rho_{e-ph}$, although the difference is small than 3% at all temperatures.

**Appendix III:** Model and experimental data of resistivity and thermal conductivity of α-U.

**Table A2.** Calculation results of α-U resistivity from 43K to 933K. Both single- and poly-crystalline results are listed.

| T (K) | Resistivity ($\times 10^{-7} \Omega$m) | | | | | | | | |
|---|---|---|---|---|---|---|---|---|---|
| | Calculations | | | | Experiments | | | | Error[2] |
| | [100] | [010] | [001] | Total | Arajs[1] | Eriksen[1] | Bennett[1] | Avg[1] | |
| 43 | 0.55 | 0.35 | 0.46 | 0.45 | 0.46 | - | - | 0.46 | -0.01 |
| 100 | 1.49 | 0.97 | 1.29 | 1.23 | 1.08 | - | - | 1.08 | 0.15 |
| 150 | 2.11 | 1.43 | 1.88 | 1.78 | 1.57 | - | - | 1.57 | 0.21 |
| 200 | 2.61 | 1.84 | 2.41 | 2.26 | 2.02 | - | - | 2.02 | 0.24 |
| 250 | 3.03 | 2.22 | 2.87 | 2.68 | 2.46 | - | - | 2.46 | 0.22 |
| 300 | 3.39 | 2.57 | 3.30 | 3.06 | 2.86 | - | - | 2.86 | 0.20 |
| 350 | 3.70 | 2.89 | 3.68 | 3.40 | 3.21 | - | 3.17 | 3.19 | 0.22 |
| 400 | 3.97 | 3.19 | 4.03 | 3.71 | 3.57 | 4.11 | 3.44 | 3.71 | -0.05 |
| 450 | 4.21 | 3.47 | 4.35 | 3.99 | 3.88 | 4.33 | 3.72 | 3.98 | -0.02 |
| 500 | 4.42 | 3.73 | 4.64 | 4.24 | 4.19 | 4.62 | 4.00 | 4.27 | -0.05 |
| 550 | 4.61 | 3.97 | 4.91 | 4.48 | 4.47 | 4.88 | 4.27 | 4.54 | -0.09 |
| 600 | 4.77 | 4.19 | 5.17 | 4.69 | 4.72 | 5.14 | 4.50 | 4.79 | -0.12 |
| 650 | 4.93 | 4.40 | 5.40 | 4.89 | 4.95 | 5.47 | 4.68 | 5.03 | -0.17 |
| 700 | 5.06 | 4.60 | 5.61 | 5.08 | 5.15 | 5.62 | 4.86 | 5.21 | -0.15 |
| 750 | 5.19 | 4.78 | 5.81 | 5.25 | 5.36 | 5.77 | 5.04 | 5.39 | -0.15 |
| 800 | 5.30 | 4.95 | 6.00 | 5.40 | 5.51 | 5.78 | 5.20 | 5.49 | -0.08 |
| 850 | 5.41 | 5.12 | 6.17 | 5.55 | 5.65 | 5.91 | 5.34 | 5.63 | -0.08 |
| 900 | 5.50 | 5.27 | 6.34 | 5.69 | - | - | - | - | - |
| 933 | 5.56 | 5.36 | 6.44 | 5.77 | - | - | - | - | - |
| The average (Mean Error (ME)) and standard deviation of error | | | | | | | | | 0.02±0.04 |
| The Root Mean Square Error (RMSE) | | | | | | | | | 0.15 |

1 Experimental resistivity values are calculated using linear interpolation of experimental data from Arajs et al. [18,51], Eriksen et al. [50], and Bennett [45]. Average experimental resistivity values are the mean values of interpolated values. All original data is given in Table A4.

2 The error is defined as $\Delta\rho_{error} = \rho_{total} - \rho_{exp(avg)}$.

**Table A3.** Calculation results of α-U thermal conductivity from 43K to 933K. Both single- and poly-crystalline results are listed.

| T (K) | Thermal conductivity (W/m-K) | | | | | |
|---|---|---|---|---|---|---|
| | Calculations | | | | Exp (Avg)[1] | Error[2] |
| | [100] | [010] | [001] | Total | | |
| 43 | 47.92 | 34.60 | 28.37 | 36.12 | - | - |
| 54[3] | 40.01 | 31.37 | 25.44 | 31.73 | 32.31 | -0.58 |
| 100 | 29.09 | 27.70 | 22.09 | 26.11 | - | - |
| 150 | 26.35 | 27.53 | 21.86 | 25.12 | - | - |
| 200 | 25.73 | 28.06 | 22.30 | 25.25 | - | - |
| 250 | 25.94 | 28.82 | 22.97 | 25.80 | - | - |
| 300 | 26.56 | 29.68 | 23.75 | 26.56 | 26.19 | 0.37 |
| 350 | 27.43 | 30.60 | 24.60 | 27.43 | 27.02 | 0.41 |
| 400 | 28.45 | 31.56 | 25.49 | 28.39 | 28.05 | 0.34 |
| 450 | 29.58 | 32.54 | 26.41 | 29.40 | 29.02 | 0.38 |
| 500 | 30.78 | 33.55 | 27.35 | 30.45 | 30.04 | 0.41 |
| 550 | 32.04 | 34.58 | 28.31 | 31.54 | 31.03 | 0.51 |

| 600 | 33.35 | 35.63 | 29.28 | 32.65 | 32.20 | 0.45 |
|---|---|---|---|---|---|---|
| 650 | 34.70 | 36.69 | 30.27 | 33.77 | 33.55 | 0.22 |
| 700 | 36.07 | 37.77 | 31.26 | 34.92 | 34.92 | 0.00 |
| 750 | 37.47 | 38.86 | 32.27 | 36.08 | 36.18 | -0.10 |
| 800 | 38.88 | 39.97 | 33.29 | 37.26 | 37.68 | -0.42 |
| 850 | 40.32 | 41.09 | 34.31 | 38.45 | 38.83 | -0.38 |
| 900 | 41.77 | 42.21 | 35.35 | 39.65 | 39.95 | -0.30 |
| 933 | 42.73 | 42.96 | 36.03 | 40.44 | 41.13 | -0.69 |
| The average (Mean Error (ME)) and standard deviation of error | | | | | | 0.09±0.11 |
| The Root Mean Square Error (RMSE) | | | | | | 0.41 |

1 Experimental thermal conductivity values are calculated using linear interpolation of experimental data from Hall et al. [12] for T≤100K, and from Ref [10,11,13,50,53,54] for T≥300K. The results of linear interpolation are not listed, but average values are given. All original data is given in Table A5. We exclude Tyler et al.'s data [10] and Hall et al's data [12] as discussed in Section 3.3.

2 The error is defined as $\Delta\kappa_{error} = \kappa_{total} - \kappa_{exp(avg)}$.

3 This data point (κ=32.31 W/m-K at 54K from Hall et al's [12]) is used to estimate the proper scale of phonon-defect scattering and to fit the parameter $A$ in Eq. (5).

**Appendix IV:** All experimental data of resistivity and thermal conductivity of α-U.

**Table A4.** Experimental α-U resistivity data.

| T (K) | R $(10^{-8}\Omega m)$ | T (K) | R $(10^{-8}\Omega m)$ | T (K) | R $(10^{-8}\Omega m)$ | T (K) | R $(10^{-8}\Omega m)$ |
|---|---|---|---|---|---|---|---|
| Raetsky[1] [100] | | Raetsky[1] [010] cont. | | Pascal et al.[1] [001] cont. | | Arajs et al.[3] cont. | |
| 67 | 9.2 | 97 | 9.8 | 200 | 24.4 | 45 | 4.6 |
| 69 | 9.7 | 99 | 10.0 | 250 | 28.8 | 51 | 5.2 |
| 70 | 10.0 | 108 | 10.9 | 288 | 32.3 | 58 | 5.8 |
| 73 | 10.4 | 113 | 11.3 | 369 | 38.1 | 65 | 6.5 |
| 73 | 10.6 | 125 | 12.4 | 472 | 45.2 | 77 | 7.8 |
| 76 | 11.2 | 150 | 14.1 | 569 | 51.0 | 88 | 8.9 |
| 85 | 12.8 | 191 | 17.3 | 672 | 55.7 | 97 | 9.8 |
| 86 | 13.0 | 201 | 18.2 | 772 | 59.2 | 107 | 10.8 |
| 87 | 13.5 | 239 | 21.0 | 871 | 61.2 | 116 | 11.6 |
| 89 | 13.7 | 263 | 22.9 | | | 127 | 12.7 |
| 95 | 14.0 | 293 | 24.8 | Eriksen et al.[2] | | 137 | 13.6 |
| 95 | 14.5 | | | 369 | 39.5 | 147 | 14.4 |
| 98 | 15.0 | Raetsky[1] [001] | | 409 | 41.6 | 157 | 15.3 |

| T (K) | K (W/m-K) | T (K) | K (W/m-K) | T (K) | K (W/m-K) | T (K) | K (W/m-K) |
|---|---|---|---|---|---|---|---|
| 101 | 15.3 | 80 | 9.7 | 458 | 43.8 | 168 | 16.3 |
| 104 | 15.7 | 84 | 10.3 | 463 | 44.3 | 177 | 17.1 |
| 107 | 16.1 | 84 | 10.5 | 504 | 46.3 | 188 | 18.1 |
| 109 | 16.5 | 91 | 11.5 | 546 | 48.6 | 198 | 18.8 |
| 113 | 17.1 | 93 | 11.9 | 585 | 50.7 | 209 | 19.7 |
| 120 | 17.8 | 109 | 13.9 | 592 | 51.1 | 219 | 20.5 |
| 121 | 18.0 | 114 | 14.6 | 623 | 52.4 | 229 | 21.3 |
| 126 | 18.3 | 128 | 16.0 | 633 | 54.2 | 241 | 22.2 |
| 127 | 18.7 | 133 | 16.6 | 702 | 56.3 | 250 | 23.0 |
| 137 | 19.6 | 163 | 19.4 | 712 | 55.9 | 261 | 23.8 |
| 141 | 20.2 | 288 | 31.2 | 729 | 57.7 | 272 | 24.6 |
| 164 | 22.3 | 300 | 32.5 | 738 | 56.7 | 283 | 25.5 |
| 166 | 22.8 | Pascal et al.[1] [010] | | 746 | 57.7 | 295 | 26.4 |
| 177 | 23.8 | 99 | 10.1 | 803 | 58.1 | 304 | 27.1 |
| 178 | 23.6 | 150 | 14.5 | 870 | 59.4 | 315 | 27.8 |
| 202 | 26.4 | 200 | 18.6 | | | 326 | 30.7 |
| 206 | 26.4 | 251 | 22.4 | Bennett[2] | | 373 | 33.5 |
| 257 | 30.9 | 289 | 25.2 | 312 | 29.6 | 398 | 35.5 |
| 291 | 34.7 | 372 | 29.8 | 574 | 44.0 | 425 | 37.2 |
| | | 473 | 35.4 | 774 | 51.2 | 499 | 41.8 |
| Raetsky[1] [010] | | 571 | 40.6 | 876 | 54.1 | 543 | 44.4 |
| 69 | 6.3 | 672 | 45.2 | | | 590 | 46.8 |
| 72 | 6.6 | 772 | 49.2 | Arajs et al.[3] | | 657 | 49.8 |
| 75 | 7.0 | 869 | 51.6 | 7 | 0.8 | 679 | 50.8 |
| 77 | 7.3 | | | 14 | 0.9 | 699 | 51.4 |
| 80 | 7.8 | Pascal et al.[1] [001] | | 20 | 1.1 | 728 | 52.9 |
| 82 | 8.1 | 73 | 10.9 | 24 | 1.6 | 758 | 53.8 |
| 84 | 8.3 | 98 | 13.3 | 29 | 2.4 | 790 | 54.8 |
| 89 | 8.8 | 149 | 19.3 | 35 | 3.2 | 832 | 56.0 |
| 91 | 9.1 | | | 40 | 3.9 | 869 | 57.1 |

1 Values read from the plot of Ref. [80]
2 Values read from the plot of Ref. [52]
3 Values read from the plot of Ref. [51]

**Table A5.** Experimental α-U thermal conductivity data.

| T (K) | K (W/m-K) | T (K) | K (W/m-K) | T (K) | K (W/m-K) |
|---|---|---|---|---|---|

| Hall et al.[1] | |
| --- | --- |
| 50 | 32 |
| 60 | 32 |
| 80 | 32 |
| 100 | 33 |

| Tyler et al.[1] | |
| --- | --- |
| 66 | 19 |
| 80 | 20.5 |
| 110 | 22.5 |
| 200 | 25.5 |
| 278 | 28 |

| Eriksen et al.[1] | |
| --- | --- |
| 373 | 26 |
| 473 | 28.5 |
| 573 | 31 |
| 673 | 33 |
| 773 | 35.5 |
| 873 | 38 |

| Howl[3] | |
| --- | --- |
| 422 | 30.5 |
| 526 | 31 |
| 614 | 33.5 |
| 648 | 35.5 |
| 685 | 36 |
| 754 | 37.5 |
| 823 | 39.5 |
| 812 | 40 |

| Babbitt et al.[1] Sample1 | |
| --- | --- |
| 293 | 25 |
| 360 | 26.5 |
| 473 | 28.5 |
| 573 | 31.5 |
| 680 | 34 |
| 780 | 38 |
| 880 | 45 |

| Babbitt et al.[1] Sample2 | |
| --- | --- |
| 293 | 25 |
| 360 | 26 |
| 473 | 28 |
| 540 | 29.5 |
| 573 | 30.5 |
| 680 | 33.5 |
| 780 | 37 |
| 880 | 41 |

| Kaity et al.[2] | |
| --- | --- |
| 300 | 27 |
| 373 | 28 |
| 473 | 31.5 |
| 573 | 33.5 |
| 673 | 36 |
| 773 | 38.5 |
| 873 | 40.5 |

| Deem et al.[1] | |
| --- | --- |
| 293 | 24 |
| 373 | 25.5 |

| Deem et al.[1] cont. | |
| --- | --- |
| 473 | 27.5 |
| 573 | 30 |
| 673 | 33 |
| 773 | 35.5 |
| 873 | 39 |

| Pearson et al.[1] | |
| --- | --- |
| 407 | 29 |
| 540 | 29.5 |
| 589 | 29.5 |
| 735 | 31.5 |
| 800 | 35 |
| 926 | 40 |

| Takahashi et al.[4] | |
| --- | --- |
| 300 | 27 |
| 400 | 28.5 |
| 500 | 30.5 |
| 600 | 32.5 |
| 700 | 35 |
| 800 | 38 |
| 900 | 40.5 |

| Touloukian et al.[3,5] | |
| --- | --- |
| 373 | 29.5 |
| 473 | 31 |
| 573 | 33.5 |
| 673 | 36 |
| 773 | 38.5 |
| 873 | 41 |

1 Values read from Ref. [46]
2 Values read from Ref. [13]
3 Values read from Ref. [10]
4 Values read from Ref. [11]
5 Recommended data

**Appendix V:** Model and experimental data of thermal conductivity of U-Zr and U-Mo.

**Table A6.** Calculation results of U-Zr thermal conductivity(W/m-K) from 373K to 873K.

| T (K) | U 4at%Zr | | | U 12at%Zr | | |
|---|---|---|---|---|---|---|
| | Calc | Touloukian[1] | Error | Calc | Touloukian[1] | Error |
| 323 | 24.6 | 23.2 | 1.4 | 21.0 | 19.8 | 1.2 |
| 373 | 25.6 | 24.1 | 1.5 | 22.1 | 21.0 | 1.1 |
| 423 | 26.6 | 25.1 | 1.5 | 23.2 | 22.0 | 1.2 |
| 473 | 27.7 | 26.1 | 1.6 | 24.3 | 22.9 | 1.4 |
| 523 | 28.8 | 27.3 | 1.5 | 25.4 | 24.0 | 1.4 |
| 573 | 29.9 | 28.6 | 1.3 | 26.5 | 25.1 | 1.4 |
| 623 | 31.0 | 29.8 | 1.2 | 27.6 | 26.6 | 1.0 |
| 673 | 32.2 | 31.1 | 1.1 | 28.7 | 28.1 | 0.6 |
| 723 | 33.3 | 32.6 | 0.8 | 29.8 | 29.5 | 0.3 |
| 773 | 34.5 | 34.1 | 0.4 | 30.9 | 30.9 | 0.1 |
| 823 | 35.7 | 35.6 | 0.8 | 32.1 | 32.4 | 0.3 |
| 873 | 36.8 | 37.2 | -0.3 | 33.2 | 34.0 | -0.8 |

| T (K) | U 14at%Zr | | | | | U 22at%Zr | | |
|---|---|---|---|---|---|---|---|---|
| | Calc | Takahashi[1] | Kaity[1] | Exp(Avg)[1] | Error | Calc | Cheon[1] | Error |
| 323 | 20.4 | 19.7 | 22.3 | 21.0 | -0.6 | 18.6 | 18.6 | 0.0 |
| 373 | 21.5 | 21.0 | 23.3 | 22.1 | -0.6 | 19.6 | 19.1 | 0.5 |
| 423 | 22.6 | 22.2 | 24.4 | 23.3 | -0.7 | 20.7 | 20.2 | 0.5 |
| 473 | 23.7 | 23.4 | 25.5 | 24.5 | -0.8 | 21.8 | 21.2 | 0.5 |
| 523 | 24.8 | 25.0 | 26.9 | 26.0 | -1.2 | 22.8 | 22.5 | 0.4 |
| 573 | 25.9 | 26.9 | 28.3 | 27.6 | -1.7 | 23.9 | 23.7 | 0.2 |
| 623 | 27.0 | 28.1 | 29.6 | 28.9 | -1.9 | 25.0 | 24.7 | 0.3 |
| 673 | 28.1 | 29.9 | 30.9 | 30.4 | -2.3 | 26.1 | 25.7 | 0.4 |
| 723 | 29.2 | 31.2 | 32.0 | 31.6 | -2.4 | 27.1 | 26.6 | 0.5 |
| 773 | 30.3 | 32.6 | 33.1 | 32.8 | -2.5 | 28.2 | 27.6 | 0.6 |
| 823 | 31.4 | 33.9 | 34.0 | 33.9 | -2.5 | 29.3 | 28.6 | 1.2 |
| 873 | 32.5 | 35.6 | - | 35.6 | -3.1 | 30.4 | 29.7 | 0.7 |

| | |
|---|---|
| The average (Mean Error (ME)) and standard deviation of error (for all concentrations) | 0.15±0.18 |
| The Root Mean Square Error (RMSE) (for all concentrations) | 1.3 |

1 Experimental thermal conductivity values are calculated using linear interpolation of experimental data from Touloukian et al. [10] for U4Zr, from Takahashi et al. [11] and Kaity et al. [13] for U14Zr (and the average experimental values for U14Zr), and from Cheon et al. [58] for U22Zr. All original data is given in Table A8.

**Table A7.** Calculation results of U-Mo thermal conductivity(W/m-K) from 373K to 873K.

| T (K) | U 0.5at%Mo | | | U 4at%Mo | | | U 11.6at%Mo | | |
|---|---|---|---|---|---|---|---|---|---|
| | Calc | Francis[1] | Error | Calc | Francis[1] | Error | Calc | Touloukian[1] | Error |
| 323 | 26.5 | - | - | 23.2 | - | - | 19.0 | - | - |
| 373 | 27.4 | - | - | 24.3 | - | - | 20.0 | - | - |
| 423 | 28.4 | - | - | 25.4 | - | - | 21.1 | 22.4 | -1.3 |
| 473 | 29.5 | 27.0 | 2.5 | 26.4 | 23.2 | 3.2 | 22.2 | 23.1 | -0.9 |
| 523 | 30.5 | 28.5 | 2.1 | 27.5 | 25.3 | 2.3 | 23.3 | 23.8 | -0.6 |
| 573 | 31.6 | 29.9 | 1.7 | 28.7 | 27.3 | 1.4 | 24.3 | 24.5 | -0.2 |
| 623 | 32.8 | 31.4 | 1.4 | 29.8 | 29.3 | 0.5 | 25.4 | 25.2 | 0.2 |
| 673 | 33.9 | 32.9 | 1.0 | 30.9 | 31.4 | -0.4 | 26.5 | 26.2 | 0.3 |
| 723 | 35.1 | 34.3 | 0.7 | 32.1 | 33.4 | -1.3 | 27.6 | 27.5 | 0.1 |
| 773 | 36.2 | 35.8 | 0.4 | 33.2 | 35.4 | -2.2 | 28.7 | 28.8 | -0.2 |
| 823 | 37.4 | 37.3 | 0.1 | 34.4 | - | - | 29.8 | - | - |
| 873 | 38.6 | - | - | 35.5 | - | - | 30.8 | - | - |

| T (K) | U 20at%Mo | | | U 21.6at%Mo | | | | | |
|---|---|---|---|---|---|---|---|---|---|
| | Calc | Matsui[1] | Error | Calc | Touloukian[1] | Klein[1] | Roy[1] | Exp(Avg)[1] | Error |
| 323 | 16.4 | 15.2 | 1.2 | 16.0 | 12.7 | 12.8 | 13.0 | 12.8 | 3.2 |
| 373 | 17.4 | 16.6 | 0.8 | 17.0 | 13.8 | 14.2 | 14.5 | 14.2 | 2.9 |
| 423 | 18.4 | 18.0 | 0.4 | 18.0 | 15.6 | 14.2 | 16.1 | 15.3 | 2.8 |
| 473 | 19.4 | 19.4 | 0.0 | 19.1 | 17.3 | 14.2 | 17.6 | 16.4 | 2.7 |
| 523 | 20.5 | 20.9 | -0.4 | 20.1 | 18.7 | 15.7 | 19.3 | 17.9 | 2.2 |
| 573 | 21.5 | 22.3 | -0.8 | 21.1 | 20.1 | 17.2 | 21.1 | 19.5 | 1.6 |
| 623 | 22.5 | 23.7 | -1.2 | 22.1 | 21.7 | 18.7 | 23.4 | 21.2 | 0.9 |
| 673 | 23.5 | 25.1 | -1.6 | 23.1 | 23.3 | 20.1 | 25.7 | 23.0 | 0.0 |
| 723 | 24.5 | 26.5 | -2.0 | 24.1 | 25.3 | 21.6 | - | 23.4 | 0.7 |
| 773 | 25.5 | 27.9 | -2.4 | 25.1 | 27.2 | 23.0 | - | 25.1 | 0.0 |
| 823 | 26.6 | 29.5 | -2.9 | 26.1 | 28.7 | 24.7 | - | 26.7 | -0.6 |
| 873 | 27.6 | 31.1 | -3.5 | 27.1 | 30.1 | 26.4 | - | 28.3 | -1.1 |
| The average (Mean Error (ME)) and standard deviation of error (for all concentrations) | | | | | | | | | 0.29±0.24 |
| The Root Mean Square Error (RMSE) (for all concentrations) | | | | | | | | | 1.6 |

1 Experimental thermal conductivity values are calculated using linear interpolation of experimental data from Francis et al. [59] for U0.5Mo and U4Mo, from Touloukian et al. [10] for U11.6Mo, from Matsui et al. [61] for U20Mo, and from Ref. [10,62,63] for U21.6Mo(and the average experimental values for U21.6Mo). All original data is given in Table A9. We exclude Konobeebsky et al.'s data [60] for U19.7Mo due to its significant error near 800K.

**Appendix VI:** Experimental data of thermal conductivity of U-Zr and U-Mo alloys.

**Table A8.** Experimental thermal conductivity data of U-Zr alloys in at%.

| T (K) | K (W/m-K) | T (K) | K (W/m-K) | T (K) | K (W/m-K) |
|---|---|---|---|---|---|
| Touloukian et al. U-4Zr[1] | | Takahashi et al. U-14Zr[2] | | Kaity et al. U-14Zr[3] | |
| 293 | 22.6 | 300 | 18.8 | 300 | 21.8 |
| 373 | 24.1 | 350 | 20.4 | 373 | 23.3 |
| 473 | 26.1 | 400 | 21.6 | 473 | 25.6 |
| 573 | 28.6 | 450 | 23.0 | 573 | 28.4 |
| 673 | 31.1 | 500 | 24.0 | 673 | 31.0 |
| 773 | 34.1 | 550 | 26.5 | 773 | 33.1 |
| 873 | 37.2 | 600 | 27.4 | 823 | 34.0 |
| | | 650 | 29.2 | | |
| Touloukian et al. U-12Zr[1] | | 700 | 30.9 | Cheon et al. U-22Zr[4] | |
| 293 | 19.2 | 750 | 31.6 | 293 | 18.2 |
| 373 | 21.0 | 800 | 33.5 | 373 | 19.1 |
| 473 | 23.0 | 850 | 34.2 | 473 | 21.2 |
| 573 | 25.1 | 880 | 36.1 | 573 | 23.7 |
| 673 | 28.2 | | | 673 | 25.7 |
| 773 | 30.9 | | | 773 | 27.5 |
| 873 | 34.0 | | | 873 | 29.7 |

1 Values read from Ref. [10]
2 Values read from the plot of Ref. [11]
3 Values read from the plot of Ref. [13]
4 Values read from the plot of Ref. [58]

**Table A9.** Experimental thermal conductivity data of U-Mo alloys in at%.

| T (K) | K (W/m-K) | T (K) | K (W/m-K) | T (K) | K (W/m-K) |
|---|---|---|---|---|---|
| Francis et al. U-0.5Mo[1] | | Touloukian et al. U-11.6Mo[2] cont. | | Matsui et al. U-20Mo[3] | |
| 473 | 27.0 | 750 | 28.2 | 293 | 14.3 |
| 573 | 29.9 | 800 | 29.5 | 373 | 16.6 |
| 673 | 32.9 | | | 473 | 19.4 |
| 773 | 35.8 | Touloukian et al. U-21.6Mo[2] | | 573 | 22.3 |
| 823 | 37.3 | 293 | 12.1 | 673 | 25.1 |
| | | 373 | 13.8 | 773 | 27.9 |
| Francis et al. U-4Mo[1] | | 473 | 17.3 | 873 | 31.1 |
| 473 | 23.2 | 573 | 20.1 | | |
| 573 | 27.3 | 673 | 23.3 | Konobeebsky et al. U-19.7Mo[3] | |
| 673 | 31.4 | 773 | 27.2 | 373 | 16.7 |
| 773 | 35.4 | 873 | 30.1 | 473 | 20.9 |
| | | | | 573 | 26.8 |
| Touloukian et al. U-11.6Mo[2] | | Klein et al. U-21.6Mo[3] | | 673 | 32.6 |
| 400 | 22.1 | 296 | 12.1 | 773 | 38.5 |
| 450 | 22.8 | 373 | 14.2 | | |
| 500 | 23.5 | 473 | 14.2 | Roy et al. U-21.6Mo[3] | |
| 550 | 24.2 | 573 | 17.2 | 323 | 12.97 |
| 600 | 24.9 | 673 | 20.1 | 485 | 17.99 |
| 650 | 25.6 | 773 | 23 | 581 | 21.34 |
| 700 | 26.9 | 873 | 26.4 | 677 | 25.94 |

1 Values read from Ref. [59]
2 Values read from Ref. [10]
3 Values read from Ref. [57]

**Appendix VII**: Alloying concentration extensibility of our model: fitting our model with one alloy data point.

For U-alloys, in our approach we have only one fitting parameter for each alloy, therefore, we can obtain the fit using only a single experimental data point. Here, we use one experimental U-Zr alloy data point to predict the thermal conductivity of U-Zr alloy to demonstrate the accuracy of our model when little alloy data is available. One example of the calculated thermal conductivity results is shown in Figure A1, where the black dot is the only experimental data point to which our model was fitted.

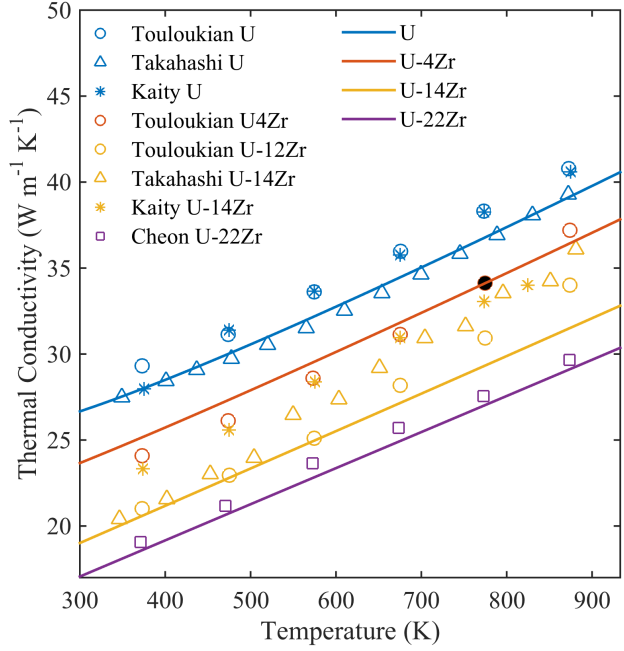

Figure A1. Calculation results of U-rich U-Zr alloys thermal conductivity from 300K to 933K (solid curves), compared with U-Zr experimental thermal conductivity data (symbols). The black dot is the data point to which the model was fitted.

The RMSE for the calculation results shown in Figure A2 is 1.8 W/m-K. By comparison, the RMSE in Section 3.4 (fitting to all data) is 1.3 W/m-K. This one-point fitting demonstrates we obtain the correct temperature and concentration dependence, and good agreement with experimental data, with just a one point fit. By individually using each of the 48 experimental data points in Figure A2 to produce the one-point fitting, the mean value of the 48 fitted RMSEs is 2.1 W/m-K, which is ~10% of the total thermal conductivity value.

**Appendix VIII:** The estimation of the impact of thermal expansion on phonon thermal conductivity.

The temperature dependence of thermal conductivity of phonon-phonon scattering at temperature higher than Debye temperature can be estimated by the Leibfried-Schlemaan formula [74]:

$$\kappa_{ph} \sim \frac{a\theta_D^3}{\gamma^2 T}, \tag{A5}$$

which leads to the following expression by differentiating with respect to volume at constant temperature:

$$\frac{\Delta\kappa_{ph}}{\kappa_{ph}} = -\left(3\gamma + 2q - \frac{1}{3}\right)\frac{\Delta V}{V}, \tag{A6}$$

where $a$ is atomic distance, $\theta_D$ is Debye temperature, $\gamma = -\left(\frac{\partial ln\theta_D}{\partial lnV}\right)_T$ is the Gruneisen parameter and $q = -\left(\frac{\partial ln\gamma}{\partial lnV}\right)_T$, where $q \approx 1$ [81,82]. For U, $\gamma = 1.7$ [75] and $\frac{\Delta V}{V} \approx 2\%$ near 600K [83], thus the phonon thermal conductivity decrease is ~13% for 600K. This leads to the total thermal conductivity to decrease <1%.

**Appendix IX**: Temperature extensibility of our model: fitting our model with resistivity data from a limited temperature range.

In the present work, the resistivity data used in our fit is all single-crystal data from 43-933K (except the data for the [100] direction, which is only from 43-300K). To test the ability of our model to extrapolate to higher temperatures, here we only use the resistivity data from 43-300K for all three directions as fitting data. All the other processes and error analysis are the same as in Section 3.3. In Figure A2, the solid black curve is the calculation result of thermal conductivity with resistivity data from 43-300K, and the dashed black curve is the original result from Section 3.3. Compared to the experimental data from 300-933K, the ME and RMSE are both 2.4 W/m-K, which is ~7% of the total thermal conductivity value.

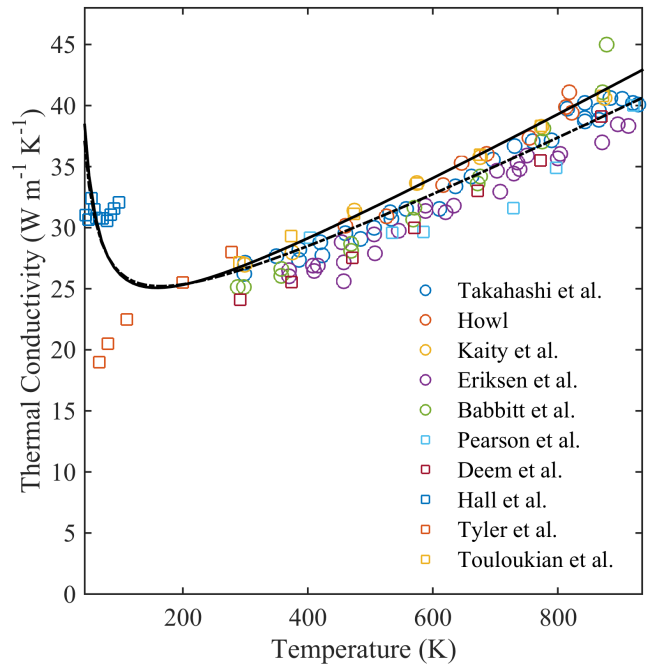

Figure A2. Calculation results of α-U thermal conductivity from 43K to 933K (curves), compared with polycrystalline α-U experimental thermal conductivity data (symbols). The solid curve is obtained by fitting to the single-crystal resistivity data from 43-300K, while the dashed curve is fitted to all available single-crystal resistivity data from 43-933K.